\documentclass[twocolumn,prd,a4paper,final,superscriptaddress,longbibliography]{revtex4-1}

\usepackage{amsmath,amssymb,amsfonts}
\usepackage[dvipdfmx]{graphicx}
\usepackage[dvipdfmx]{hyperref}

\usepackage{bbold}
\usepackage[T1]{fontenc}
\usepackage{epsfig}
\usepackage{epstopdf}
\usepackage{bbold}
\usepackage{comment}
\usepackage{bm}

\usepackage{natbib}

\begin{document}
	
	\title{Compact, charged boson-stars, -shells in the $\mathbb{C}P^N$ gravitating nonlinear sigma model}
	
	\author{Nobuyuki Sawado}
	\email{sawadoph@rs.tus.ac.jp}
	\affiliation{Department of Physics, Tokyo University of Science, Noda, Chiba 278-8510, Japan}

	\author{Shota Yanai}
	\email{phyana0513@gmail.com}
	\affiliation{Department of Physics, Tokyo University of Science, Noda, Chiba 278-8510, Japan}

	\vspace{.5 in}
	\small

	\date{\today}
	
	\begin{abstract}
	We study $U(1)$ gauged gravitating compact $Q$-ball, $Q$-shell solutions in a nonlinear sigma model with the target space $\mathbb{C}P^N$. 
	The models with odd integer $N$ and a special potential can be parameterized by $N$-th complex scalar fields and they 
	support compact solutions.  
	Implementing the $U(1)$ gauge field in the model, the behavior of the solutions become complicated than the global model.
	Especially, they exhibit branch, i.e., two independent solutions with same shooting parameter. 
	The energy of the solutions in the first branch behaves as $E\sim Q^{5/6}$ for small $Q$, 
	where $Q$ stands for the $U(1)$ Noether charge. 
	For the large $Q$, it gradually deviates from the scaling $E\sim Q^{5/6}$ and, for the $Q$-shells it is $E\sim Q^{7/6}$, 
	which forms the second branch.  
	A coupling with gravity allows for harboring of the Schwarzschild black holes for the $Q$-shell 
	solutions, forming the charged boson shells. The space-time then consist of a charged black hole in the interior of the shell, 
	surrounded by a $Q$-shell, and the outside becomes a Reissner-Nordstr\"om space-time. 
	These solutions inherit the scaling behavior of the flat space-time. 
	\end{abstract}
	
	%:Pacs & Keywords

	\maketitle 
	\section{Introduction}
	\label{Intro}
	
	A complex scalar field theory with some self-interactions has stationary
	soliton solutions called $Q$-balls~\cite{Friedberg:1976me,Coleman:1985ki,Leese:1991hr,Loiko:2018mhb}. 
	$Q$-balls have attracted much attention in the studies of 
	evolution of the early Universe~\cite{Friedberg:1986tq,Lee:1986ts}.
	In supersymmetric extensions of the standard model, $Q$-balls appear as
	the scalar superpartners of baryons or leptons forming coherent states with 
	baryon or lepton number. They may survive as a major ingredient of dark matter
	~\cite{Kusenko:1997zq,Kusenko:1997si,Kusenko:1997vp}.
	The $U(1)$ invariance of the scalar field 
	leads to the conserved charge $Q$ which, if the theory is coupled with the 
	electromagnetism it identifies as the electric charge of the constituents.
	
	Our analysis is based on the model defined in 3+1 dimensions and it has the Lagrangian density
	\begin{align}
	{\cal L}=-\frac{M^2}{2}\mathrm{Tr}(X^{-1}\partial_\mu X)^2-\mu^2V(X),
	\label{lag0}
	\end{align}
	where the `V-shaped' potential
	\begin{align}
	V(X)=\frac{1}{2}[\mathrm{Tr}(I-X)]^{1/2}
	\end{align} 
	is employed for constructing the compact solutions. The behavior of fields at the outer border of compacton implies $X\to I$. 
	The coupling constants $M$, $\mu$ have dimensions of $(\mathrm{length})^{-1}$ and $(\mathrm{length})^{-2}$, respectively. 
	The principal variable $X$ successfully parameterizes the coset space
	 $SU(N+1)/U(N)\sim \mathbb{C}P^N$.
	The principal variable parametrized by complex fields $u_i$ takes the form
	\begin{align}
	X(g)=
	\left(\begin{array}{cc}
	I_{N\times N} & 0 \\
	0 & -1 
	\end{array}\right)+
	\frac{2}{\vartheta^2}
	\left(\begin{array}{cc}
	-u\otimes u^\dagger & iu \\
	iu^\dagger & 1  
	\end{array}\right)
	\label{principalu}
	\end{align}
	where $\vartheta:=\sqrt{1+u^\dagger\cdot u}$. 		
	Thus the $\mathbb{C}P^N$ Lagrangian of the model (\ref{action}) 
	takes the form
	\begin{align}
	{\cal L}_{\mathbb{C}P^N}=-M^2g^{\mu\nu}\tilde{\tau}_{\nu\mu}-\mu^2 V
	\label{lagrangianu}
	\end{align}
	where
	\begin{align}
	\tilde{\tau}_{\nu\mu}=-\frac{4}{\vartheta^4}\partial_\mu u^\dagger\cdot \Delta^2\cdot \partial_\nu u,~~
	\Delta_{ij}^2:=\vartheta^2\delta_{ij}-u_iu_j^*\,.
	\end{align}
	The model possesses the compactons~\cite{Klimas:2017eft} and also the compact boson stars~\cite{Klimas:2018ywv}.
	Compactons are field configurations that exist on finite size supports. Outside this support, the field is 
	identically zero. For example, the signum-Gordon model; i.e., the scalar field model with standard
	kinetic terms and V-shaped potential gives rise to such solutions~\cite{Arodz:2008jk,Arodz:2008nm}. 
	Interestingly, when the scalar field is coupled with electromagnetism, the structure changes drastically. 
	
	Maxwell gauged solitons of non-linear sigma model have been studied in many years. The gauged O(3) 
	model~\cite{Schroers:1995he}, the baby-skyrmions~\cite{Gladikowski:1995sc,Navarro-Lerida:2018siw} 
	and the magnetic skyrmions~\cite{Schroers:2019hhe}, 
	$\mathbb{C}P^1$ model~\cite{Tchrakian:1995np} are well-known examples. 
	For $\mathbb{C}P^N,~N>1$, there is a work for a $\mathbb{C}P^2$ of Maxwell gauged model~\cite{Loginov:2016yeh}.
	
	$Q$-balls resulting from local $U(1)$ symmetry are studied in the literature~\cite{Lee:1988ag,Anagnostopoulos:2001dh,Levi:2001aw,
	Gulamov:2013cra,Brihaye:2014gua,Tamaki:2014oha,Gulamov:2015fya,Loiko:2019gwk,Loginov:2020xoj}.
	Such $Q$-balls may be unstable for large values of their charge because of the repulsion mediated by the gauge force and  
	the fermions or the scalar fields with opposite charge may reduce such repulsions~\cite{Anagnostopoulos:2001dh}. 

	In the case of the compactons, when the scalar field is coupled with electromagnetism then the inner radius emerges, 
	i.e., the scalar field vanishes also in the central region $r<R_{\rm in}$. 
	Thus, the matter field exists in the region $R_{\rm in}\leqq r \leqq R_{\rm out}$. Such configurations of fields are called $Q$-shells. 
	Such shell solutions have no restrictions on upper bound for $|Q|$.
	The authors claim that the energy of compact $Q$-balls scales as $\sim Q^{5/6}$
	and of $Q$-shells for large $Q$ as $\sim Q^{7/6}$. 
	It clearly indicates that the $Q$-balls are stable against the decay while the $Q$-shells may be 
	unstable. 
	
	Boson stars are the gravitating objects of such $Q$-balls. There are large number of articles concerning the boson stars
	~\cite{Lee:1991ax, Friedberg:1986tq, Jetzer:1991jr, Kleihaus:2009kr,Kleihaus:2010ep,Liebling:2012fv}.  
	For the compact boson stars and shells of $U(1)$ gauged model are studied in \citep{Kumar:2014kna,Kumar:2015sia,Kumar:2016sxx}.
	For the boson shell configurations, one possibility is the case that the gravitating boson 
	shells surround a flat Minkowski-like interior region $r<R_{\rm in}$ while the exterior
	region $r>R_{\rm out}$ is the exterior of an Reissner-Nordstr\"om solution. 
	Another and even more interesting possibility are the existence of the charged black hole 
	in the interior region. The gravitating boson shells can harbor a black hole. 
	Since the black hole is surrounded by a shell of scalar fields, such
	fields outside of the event horizon may be interpreted as a scalar hair. 	
	Such possibility has been considered as contradiction of the no-hair conjecture.  
	The higher dimensional generalizations have been considered in \citep{Hartmann:2012da,Hartmann:2013kna}.

	The excited boson stars are very important not only of the theoretical interest and also 
	for astrophysical observations~\cite{Brihaye:2008cg,Bernal:2009zy,Collodel:2017biu,Alcubierre:2018ahf,Alcubierre:2019qnh}. 
	The multistate boson stars which is superposed ground and excited states boson star solutions are considered 
	for obtaining realistic rotation curves of spiral galaxies~\cite{Bernal:2009zy}. The authors of 
	\cite{Alcubierre:2018ahf,Alcubierre:2019qnh} proposed boson star solutions for a collection of an arbitrary odd number 
	$N$ of complex scalar fields with an internal symmetry $U(N)$. They are  new excited solutions with angular momentum $\ell$
	so is dubbed as $\ell-$boson stars. Our $\mathbb{C}P^N$ boson-stars share many common features with them. 
	
 	In this paper, we explore the $U(1)$ gauged the gravitating boson-shells. Properties of 
	the harbor type solutions are also discussed.
      Especially we examine detailed energy scaling property about the $\sim Q^{5/6}$ behavior 
	for the gravitating, or the harbor type solutions. 
	
	The paper is organized as follows. In Section II we shall describe the model,
	coupled to the gravitation. Ansatz for the parametrization of the $\mathbb{C}P^N$ field 
	is given in this section. 	
	Section III is analysis of in the flat space-time, i.e., $Q$-balls and $Q$-shells. 
	We give the gravitation solutions in Section IV. Scaling relation between the energy 
	and the charge in the boson-stars and -shells are discussed in Section V. 
	Conclusions and remarks are presented in the last Section.

%%%%%%%%%%%%%%%%%%%%%%%%%%%%%%%%%%%%%%%%%%%%%%%%%%%%%%%%%%%%%%%%%%%%%%%%%%%%%%%%%%%%%%%%%%%%%%%%%%%%%%%%%%%%%%%%%
\begin{figure*}[t]
  \begin{center}
    \includegraphics[width=80mm]{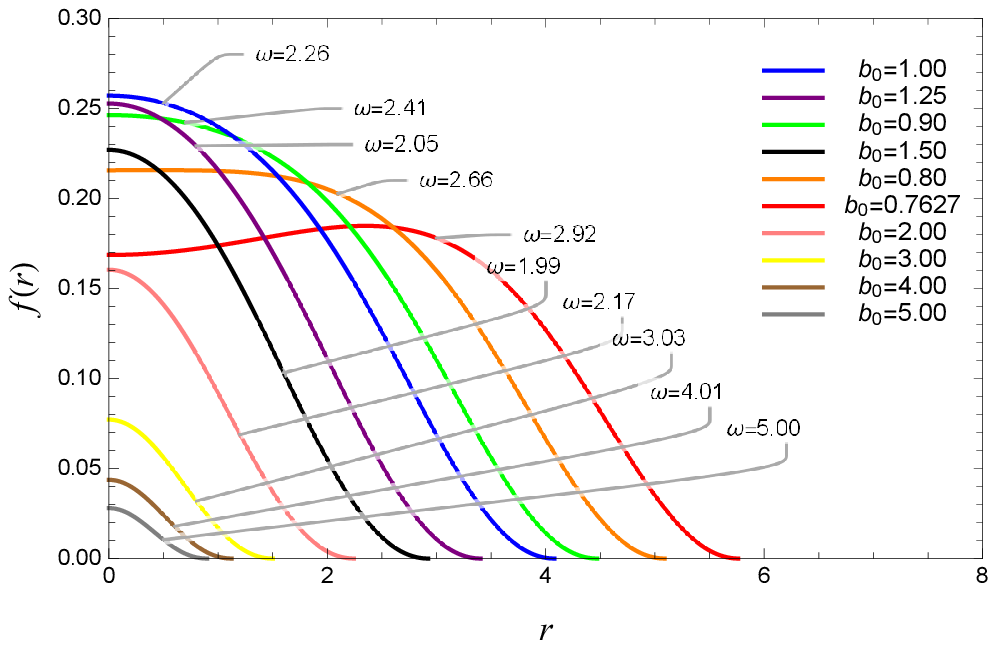}~~
    \includegraphics[width=80mm]{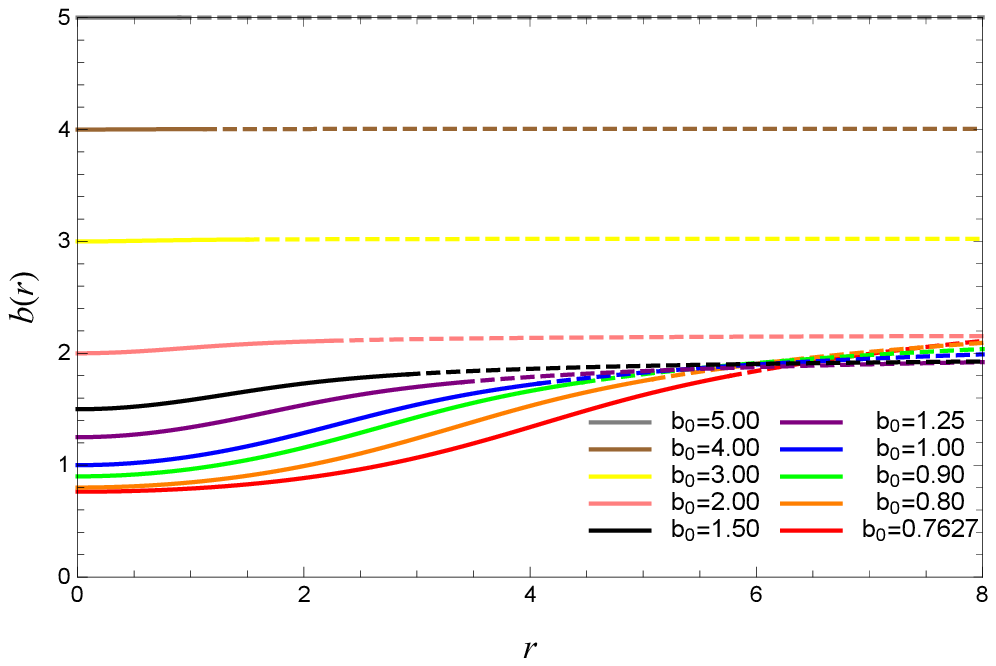}\\
    \includegraphics[width=80mm]{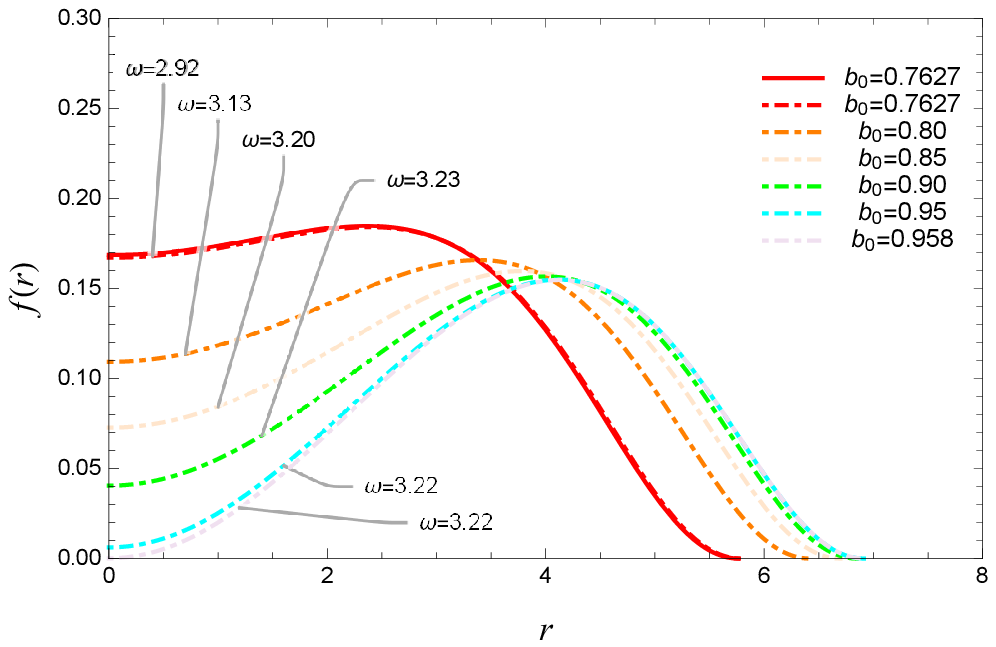}~~
    \includegraphics[width=80mm]{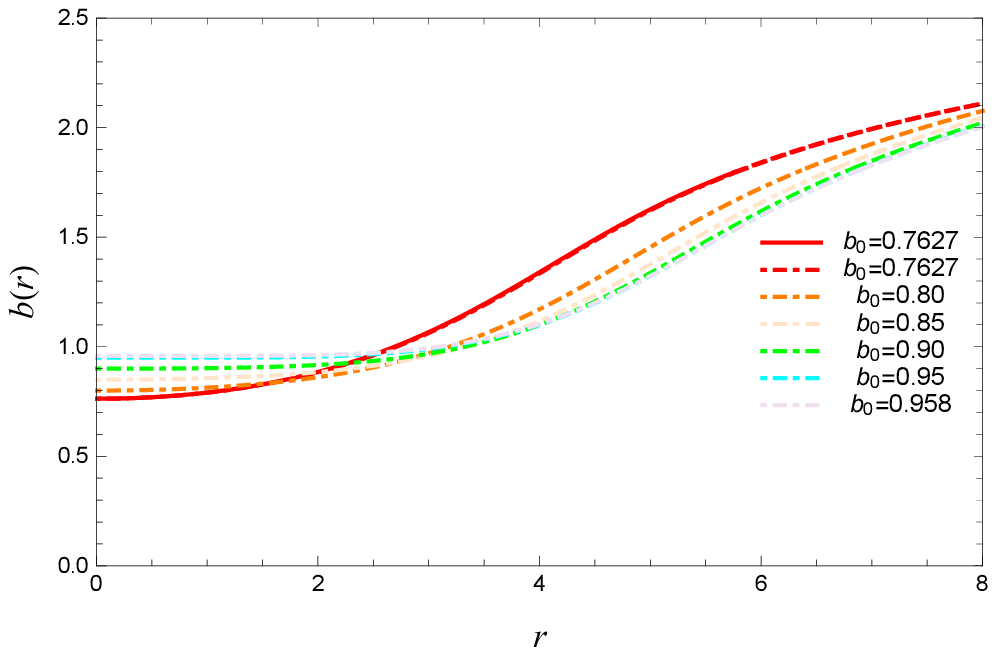}
    \vspace{5mm}

    \caption{\label{CP1} The gauged $Q$-ball solution for the $\mathbb{C}P^{1}$ case. 
{\it Top Left}:~The matter profile function $f(r)$ of the first branch. {\it Top Right}:~The gauge function $b(r)$ of the first branch. 
{\it Bottom Left}:~The matter profile function $f(r)$ of the second branch. {\it Bottom Right}:~The gauge function $b(r)$ of the second branch.
Solutions of the first branch are plotted with bold line and the second branch are plotted with dot-dashed line. 
Solutions of the vacuum equations are depicted with the dashed line. 
Distinct curves correspond with different values of shooting parameter $b_{0}$.}
  \end{center}
	\end{figure*}
%%%%%%%%%%%%%%%%%%%%%%%%%%%%%%%%%%%%%%%%%%%%%%%%%%%%%%%%%%%%%%%%%%%%%%%%%%%%%%%%%%%%%%%%%%%%%%%%%%%%%%%%%%%%%%%%%	
	
%%%%%%%%%%%%%%%%%%%%%%%%%%%%%%%%%%%%%%%%%%%%%%%%%%%%%%%%%%%%%%%%%%%%%%%%%%%%%%%%%%%%%%%%%%%%%%%%%%%%%%%%%%%%%%%%%
\begin{figure*}[t]
  \begin{center}
    \includegraphics[width=80mm]{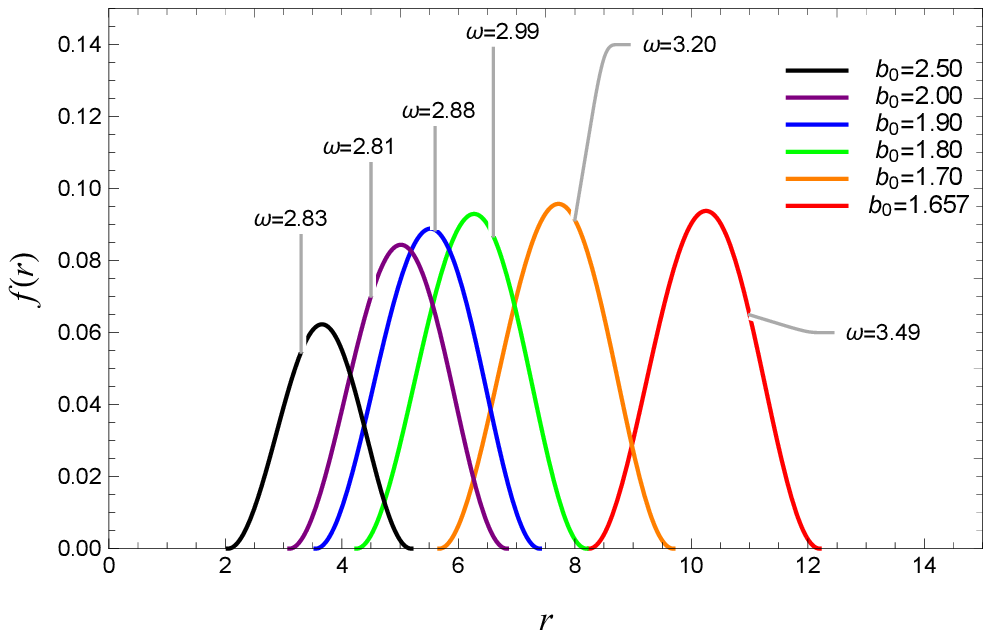}~~
    \includegraphics[width=80mm]{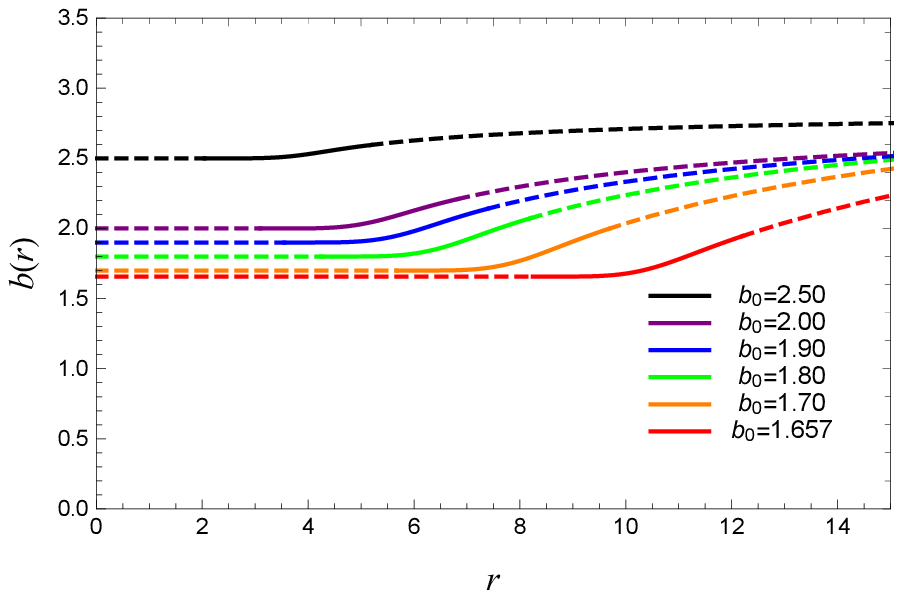}\\
	\includegraphics[width=80mm]{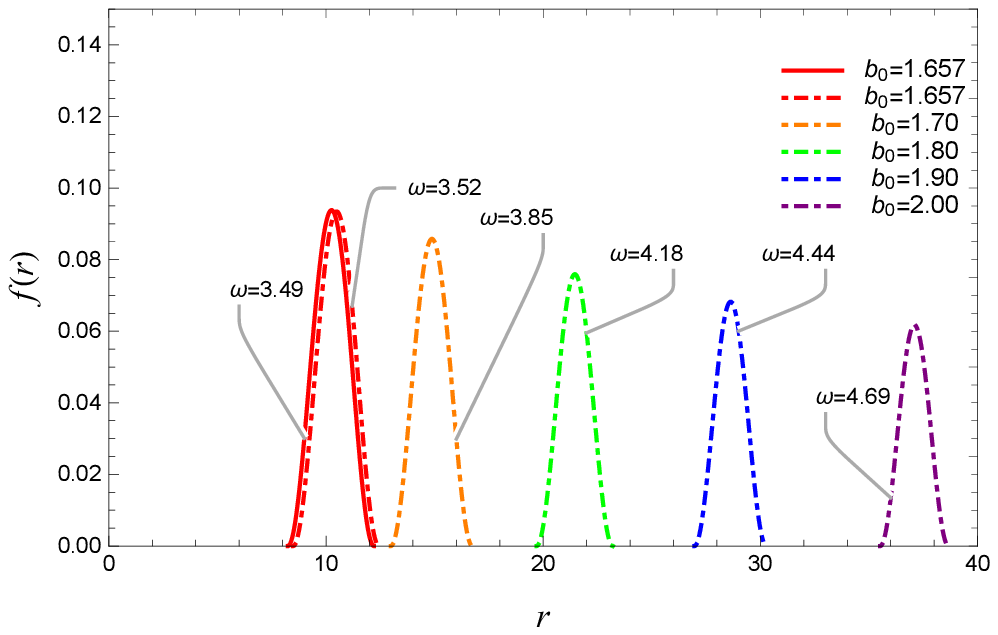}~~
    \includegraphics[width=80mm]{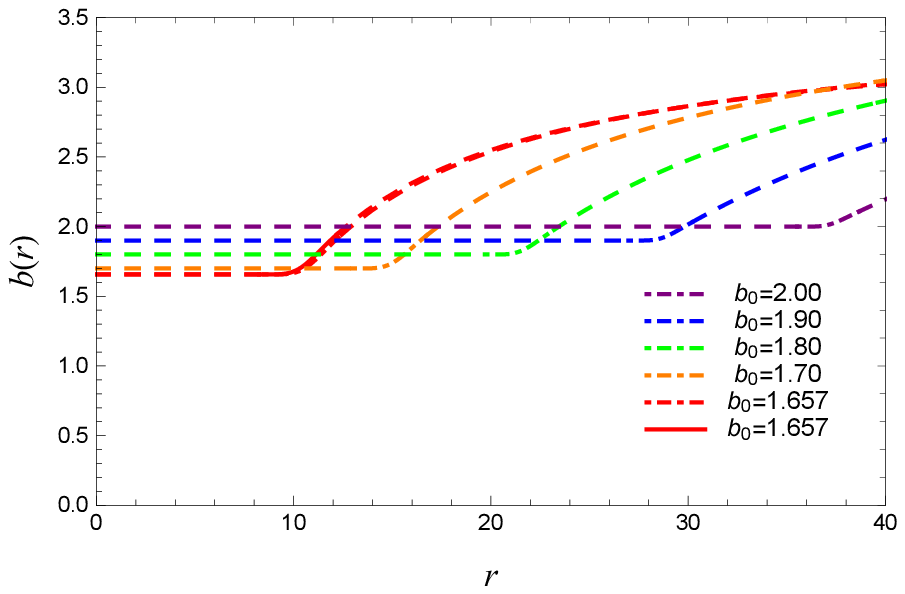}
    \vspace{5mm}
    \caption{\label{CP11} The gauged $Q$-shell solution for the $\mathbb{C}P^{11}$ case. 
{\it Top Left}:~The matter profile function $f(r)$ of the first branch. {\it Top Right}:~The gauge function $b(r)$ of the first branch. 
{\it Bottom Left}:~The matter profile function $f(r)$ of the second branch. {\it Bottom Right}:~The gauge function $b(r)$ of the second branch. 
Solutions of the first branch are plotted with bold line and the second branch are plotted with dot-dashed line. 
Solutions of the vacuum equations are depicted with the dashed line. 
Distinct curves correspond with different values of shooting parameter $b_{0}$.}
  \end{center}
	\end{figure*}
%%%%%%%%%%%%%%%%%%%%%%%%%%%%%%%%%%%%%%%%%%%%%%%%%%%%%%%%%%%%%%%%%%%%%%%%%%%%%%%%%%%%%%%%%%%%%%%%%%%%%%%%%%%%%%%%%

	\section{The model}
	\label{model}
	
	\subsection{The action, the equations of motion}
	We start with the action of self-gravitating complex fields $u_i$ coupled 
	to Einstein gravity
	\begin{align}
	&S=\int d^4x\sqrt{-g}\biggl[
	\frac{R}{16\pi G}-\frac{1}{4}g^{\mu\lambda}g^{\nu\sigma}F_{\mu\nu}F_{\lambda\sigma} \nonumber \\
	&\hspace{4cm}-M^2g^{\mu\nu}\tau_{\nu\mu}-\mu^2 V
	\biggr]\,, 
	\label{action} \\
	&\tau_{\nu\mu}=-\frac{4}{\vartheta^4}D_\mu u^\dagger\cdot \Delta^2\cdot D_\nu u,~~
	\Delta_{ij}^2:=\vartheta^2\delta_{ij}-u_iu_j^*\,.
	\end{align}
	where $G$ is Newton's gravitational constant. $F_{\mu\nu}$ is the standard electromagnetic field tensor 
	and the complex fields $u_i$ also are minimally coupled to the Abelian gauge fields $A_\mu$ through
	$D_\mu=\partial_\mu-ieA_\mu$.

	The variation of the action with respect to the metric leads to Einstein's equations
	\begin{align}
	G_{\mu\nu}=8\pi GT_{\mu\nu},\quad{\rm where}\quad G_{\mu\nu}\equiv R_{\mu\nu}-\frac{1}{2}g_{\mu\nu}R
	\label{einstein_formal}
	\end{align}
	where the stress-energy tensor reads
	\begin{align}
	T_{\mu\nu}=g_{\mu\nu}(M^2g^{\lambda\sigma}\tau_{\sigma\lambda}+\frac{1}{4}g^{\lambda\sigma}g^{\eta\delta}F_{\lambda\eta}F_{\sigma\delta}+\mu^2V)
	\nonumber \\
	-2M^2\tau_{\nu\mu}-g^{\lambda\sigma}F_{\mu\lambda}F_{\nu\sigma}\,.
	\label{stress_formal}
	\end{align}
	The field equations of the complex fields are obtained by variation of the Lagrangian with respect to $u_i^*$
	\begin{align}
	&\frac{1}{\sqrt{-g}}D_\mu (\sqrt{-g} D^\mu u_i)-\frac{2}{\vartheta^2}(u^\dagger\cdot D^\mu u)D_\mu u_i \nonumber \\
	&+\frac{\mu^2}{4M^2}\vartheta^2\sum_{k=1}^{N}\biggl[(\delta_{ik}+u_iu_k^*)\frac{\partial V}{\partial u_k^*}\biggr]=0\,.
	\label{CPNeq}
	\end{align}
	The Maxwell's equations read 
	\begin{align}
	\frac{1}{\sqrt{-g}}\partial_\nu (\sqrt{-g}F^{\nu\mu})=\frac{4ie}{\vartheta^4}M^2(u^\dagger\cdot D^\mu u-D^\mu u^\dagger\cdot u)\,.
	\label{Maxwell}
	\end{align}
	
	It is convenient to introduce the dimensionless coordinates 
	\begin{align}
	x_\mu \to \frac{\mu}{M}x_\mu
	\end{align}
	and also $A_\mu \to \mu/M A_\mu$. 
	We also restrict $N$ to be odd, i.e., $N:=2n+1$.  For solutions with vanishing magnetic field 
	the ansatz has the form
	\begin{align}
	&u_m(t,r,\theta,\varphi)=\sqrt{\frac{4\pi}{2n+1}}f(r)Y_{nm}(\theta,\varphi)e^{i\omega t}\,, 
	\label{ansatzcpn}\\
	&A_\mu(t,r,\theta,\varphi)dx^\mu=A_t(r)dt
	\label{ansatzgauge}
	\end{align}
	allows for reduction of the partial differential equations to the system of radial ordinary 
	differential equations.
	$Y_{nm}, -n\leq m \leq n$ are the standard spherical harmonics and $f(r)$ is the profile function. 
	Each $2n+1$ field $u=(u_m)=(u_{-n},u_{-n+1},\cdots,u_{n-1},u_n)$ is associated with one of $2n+1$
	spherical harmonics for given $n$. 
	The relation 
	$\sum_{m=-n}^n Y_{nm}^*(\theta,\varphi)Y_{nm}(\theta,\varphi)=\dfrac{2n+1}{4\pi}$
	is very useful for obtaining an explicit form of many inner products. 
	We introduce a new gauge field concerning the gauge field for convenience
	\begin{align}
	b(r):=\omega-eA_t(r)\,.
	\label{ansatzgaugew}
	\end{align}
	
	Using the ansatz we find the dimensionless Lagrangian of the $\mathbb{C}P^N$ model in the form 
	\begin{align}
	&\tilde{\mathcal{L}}_{\mathbb{C}P^N}
	=-\frac{\kappa}{4}g^{\mu\lambda}g^{\nu\sigma}F_{\mu\nu}F_{\lambda\sigma}-g^{\nu\mu}\tau_{\nu\mu}-V
	\nonumber \\
	&=\frac{\kappa b'^2}{2A^2e^2}+\frac{4b^2f^2}{A^2C(1+f^2)^2}-\frac{4Cf'^2}{(1+f^2)^2}-\frac{4n(n+1)f^2}{r^2(1+f^2)}-V
	\label{effectivelag}
	\end{align}  
	where we have introduced a dimensionless constant $\kappa:=\mu^2/M^4$ for convenience.
	
	There is a symmetry $b\to-b$, i.e., $\omega\to-\omega,eA_t\to -eA_t$ in (\ref{effectivelag}), so
	one can simply assume that $\omega\geqq 0$~\cite{Lee:1988ag}.
	In this paper, we shall not adopt the above symmetry and examine the case with both signs of $\omega$. 
	We shall see that our obtained solutions always satisfy $b> 0$, and the Noether charge is positive definite.   
	
	For the ansatz (\ref{ansatzcpn})-(\ref{ansatzgaugew}),  
	a suitable form of line element is the standard spherically symmetric Schwarzschild-like coordinates defined by
	\begin{align}
	ds^2 &= g_{\mu \nu}dx^{\mu}dx^{\nu} \nonumber \\
	&=A^2(r)C(r)dt^2 - \frac{1}{C(r)}dr^2 - r^2(d\theta^2 +\sin^2 \theta d\varphi^2)\,.
	\label{metric}
	\end{align}    
	Substituting (\ref{metric}) and (\ref{ansatzcpn})-(\ref{ansatzgaugew}) 
	into the Einstein field equations (\ref{einstein_formal}),
	we get their components  
	\begin{align}
	&(tt):~~\frac{[r(1-C)]'}{r^2} 
	=\alpha \Bigl[ \frac{4 b^2 f^2}{A^2C(1+f^2)^2}
	+\frac{4C f'^2}{(1+f^2)^2}
		\nonumber \\ 
	&\hspace{2.5cm}+\frac{4n(n+1)f^2}{r^2(1+f^2)}+\frac{\kappa b'^2}{2e^2A^2}+\frac{f}{\sqrt{1+f^2}}\Bigr]\,,
	\label{metrictt} 
	\\
	&(rr):~~\frac{2rCA'-A[r(1-C)]'}{r^2 A}
	=\alpha \Bigl[\frac{4 b^2 f^2}{A^2C(1+f^2)^2}
	\nonumber \\
	&\hspace{1cm}+ \frac{4Cf'^2}{(1+f^2)^2}-\frac{4n(n+1)f^2}{r^2(1+f^2)} 
	- \frac{\kappa b'^2}{2A^2  e^2}-\frac{f}{\sqrt{1+f^2}} \Bigl]\,,
	\label{metricrr}
	\\
	&(\theta\theta):~~\frac{3rA'C'+2C(A'+rA'')+A(2C'+rC'')}{2rA} \nonumber \\	
	&\hspace{0.8cm}= \alpha \Bigl[\frac{4 b^2 f^2}{A^2C(1+f^2)^2}-\frac{4Cf'^2}{(1+f^2)^2}
	+\frac{\kappa b'^2}{2A^2 e^2} - \frac{f}{\sqrt{1+f^2}}\Bigr]\,,
	\label{metricqq}
	\end{align}
	where $\alpha$ is a dimensionless coupling constant concerning to the gravity
	\begin{align}
	\alpha:=8\pi G\mu^2.
	\end{align}
	From (\ref{metrictt}),(\ref{metricrr}) one can construct the equations of motion of $A(r),C(r)$
	\begin{align}
	&A'=4\alpha r \Bigl[\frac{b^2 f^2}{A^2 C^2(1+f^2)^2}+\frac{f'^2}{(1+f^2)^2} \Bigr]\,, 
	\label{eq:N}\\
	&C'=\frac{1-C}{r}
	\nonumber \\
	&-\alpha r\Bigl[\frac{4b^2 f^2}{A^2 C(1+f^2)^2}+\frac{4Cf'^2}{(1+f^2)^2}+\frac{4n(n+1)f^2}{(1+f^2)r^2}
	\nonumber \\
	&\hspace{1cm}+\frac{\kappa b'^2}{2A^2e^2} +\frac{f}{\sqrt{1+f^2}}\Bigr]\,.
	\label{eq:C}
	\end{align}
	Plugging the ansatz (\ref{ansatzcpn})-(\ref{ansatzgaugew}) into the matter field equation (\ref{CPNeq}) and the Maxwell's equations (\ref{Maxwell}), 
	we have 
	\begin{align}
	&Cf'' +C'f'+\frac{A'Cf'}{A} +\frac{2C}{r}f' - \frac{n(n+1)f}{r^2}
	\nonumber \\
	&\hspace{1cm}+\frac{(1-f^2)b^2f}{A^2C(1+f^2)}-\frac{2Cff'^2}{(1+f^2)} -\frac{1}{8}\sqrt{1+f^2} =0\,, 
	\label{eq:f}\\
	&\kappa b''+\frac{2r'A-A'r}{Ar}\kappa b' - \frac{8e^2}{C}\frac{bf^2}{(1+f^2)^2} =0\,.
	\label{eq:b}
	\end{align}
	Thus, we solve a four coupled equations (\ref{eq:N})-(\ref{eq:b}) varying the parameters $\alpha$
	with fixed $\kappa,e$ (in this paper we simply set $\kappa=e=1$).
	
	The dimensionless Hamiltonian of the model are easily obtained 
	\begin{align}
	\mathcal{H}_{\mathbb{C}P^N}&=\frac{\partial\tilde{\mathcal{L}}_{\mathbb{C}P^N}}{\partial(\partial_0u_i^*)}\partial_0u_i^*
	+\frac{\partial\tilde{\mathcal{L}}_{\mathbb{C}P^N}}{\partial(\partial_0u_i)}\partial_0u_i - \tilde{\mathcal{L}}_{\mathbb{C}P^N}
	\nonumber \\
	&=\frac{8\omega bf^2}{A^2C(1+f^2)^2}-\frac{\kappa b'^2}{2A^2e^2}-\frac{4b^2f^2}{A^2C(1+f^2)^2}
	\nonumber \\
	&\hspace{1cm}+\frac{4Cf'^2}{(1+f^2)^2}+\frac{4n(n+1)f^2}{r^2(1+f^2)}+V\,.
	\end{align}
	The total energy is thus given by
	\begin{align}
	&E=4\pi\int r^2dr
	\biggl[
	\frac{\kappa b'^2}{2Ae^2}+\frac{4b^2f^2}{AC(1+f^2)^2}
	\nonumber \\
	&\hspace{1cm}+\frac{4ACf'^2}{(1+f^2)^2}+\frac{4An(n+1)f^2}{r^2(1+f^2)}+AV\biggr]\,.
	\label{energy}
	\end{align}

%%%%%%%%%%%%%%%%%%%%%%%%%%%%%%%%%%%%%%%%%%%%%%%%%%%%%%%%%%%%%%%%%%%%%%%%%%%%%%%%%%%%%%%%%%%%%%%%%%%%%%%%%%%%%%%%%
\begin{figure*}[t]
  \begin{center}
    \includegraphics[width=80mm]{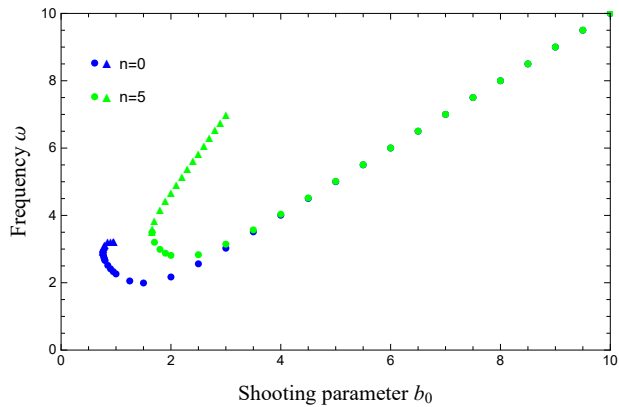}
    \caption{\label{b0w} A relation between shooting parameter $b_{0}$ and frequency $\omega$ 
for $\mathbb{C}P^{1}$ (blue) and $\mathbb{C}P^{11}$ (green).
The dots correspond to solutions with different $\omega$ of the first branch and the triangles are of the second branch. }
  \end{center}
	\end{figure*}
%%%%%%%%%%%%%%%%%%%%%%%%%%%%%%%%%%%%%%%%%%%%%%%%%%%%%%%%%%%%%%%%%%%%%%%%%%%%%%%%%%%%%%%%%%%%%%%%%%%%%%%%%%%%%%%%%
%%%%%%%%%%%%%%%%%%%%%%%%%%%%%%%%%%%%%%%%%%%%%%%%%%%%%%%%%%%%%%%%%%%%%%%%%%%%%%%%%%%%%%%%%%%%%%%%%%%%%%%%%%%%%%%%%
\begin{figure*}[t]
  \begin{center}
    \includegraphics[width=80mm]{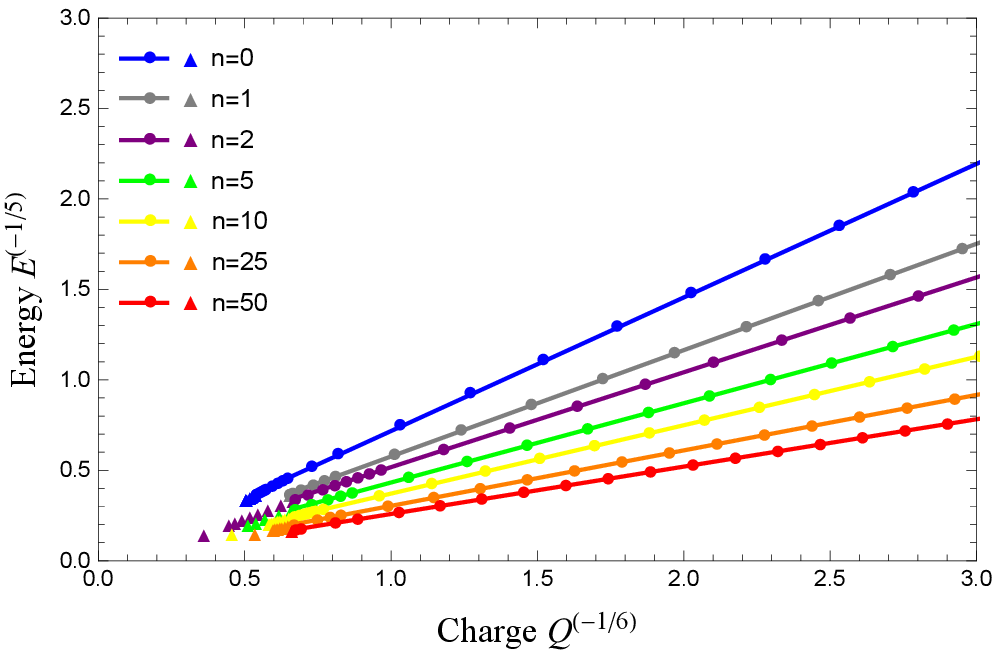}~~
    \includegraphics[width=80mm]{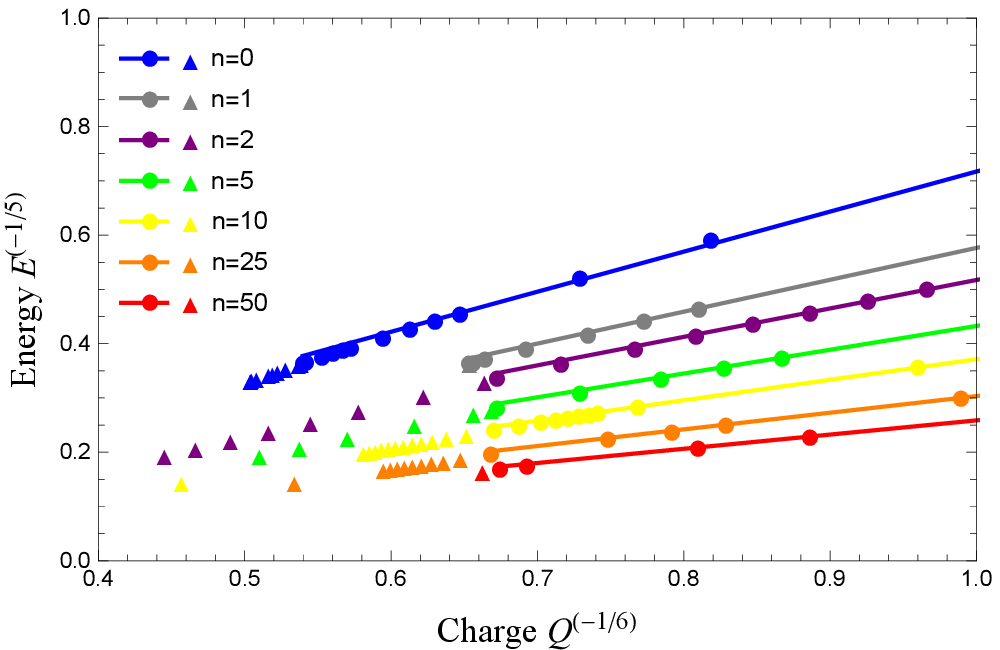}\\
    \vspace{5mm}
    \caption{\label{Q6E5f} {\it Left}:~The relation between $E^{-1/5}$ and $Q^{-1/6}$ for several gauged solutions. 
      {\it Right}: The same as the left one but the plot is enlarged in region of high $Q,E$.
	The dots correspond to solutions of first branch. The triangles are of the second branch. }
  \end{center}
	\end{figure*}
%%%%%%%%%%%%%%%%%%%%%%%%%%%%%%%%%%%%%%%%%%%%%%%%%%%%%%%%%%%%%%%%%%%%%%%%%%%%%%%%%%%%%%%%%%%%%%%%%%%%%%%%%%%%%%%%%
%%%%%%%%%%%%%%%%%%%%%%%%%%%%%%%%%%%%%%%%%%%%%%%%%%%%%%%%%%%%%%%%%%%%%%%%%%%%%%%%%%%%%%%%%%%%%%%%%%%%%%%%%%%%%%%%%
\begin{figure*}[t]
  \begin{center}
    \includegraphics[width=80mm]{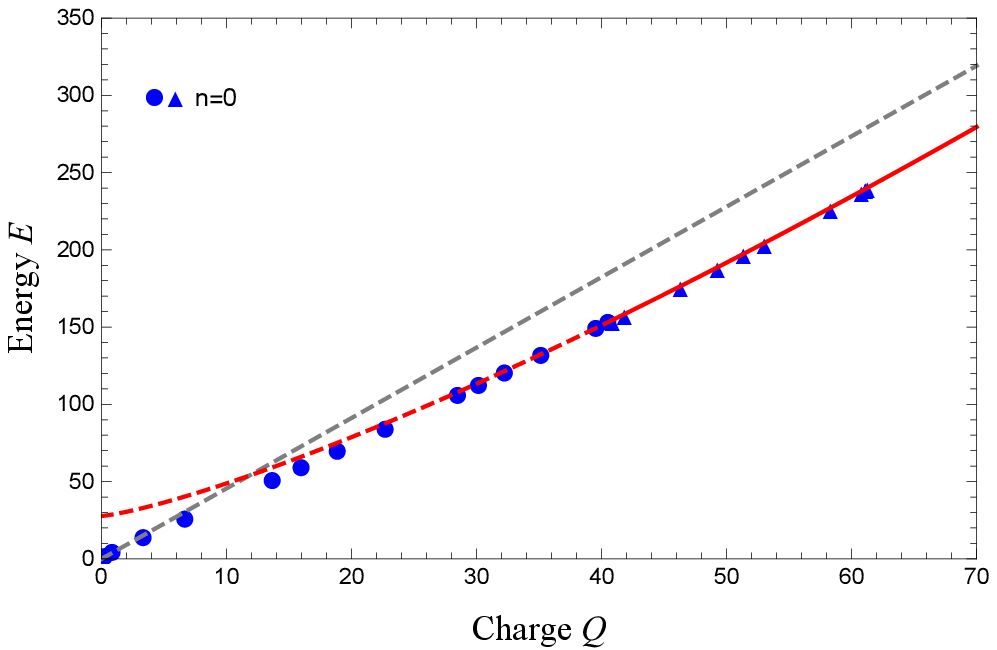}~~
    \includegraphics[width=80mm]{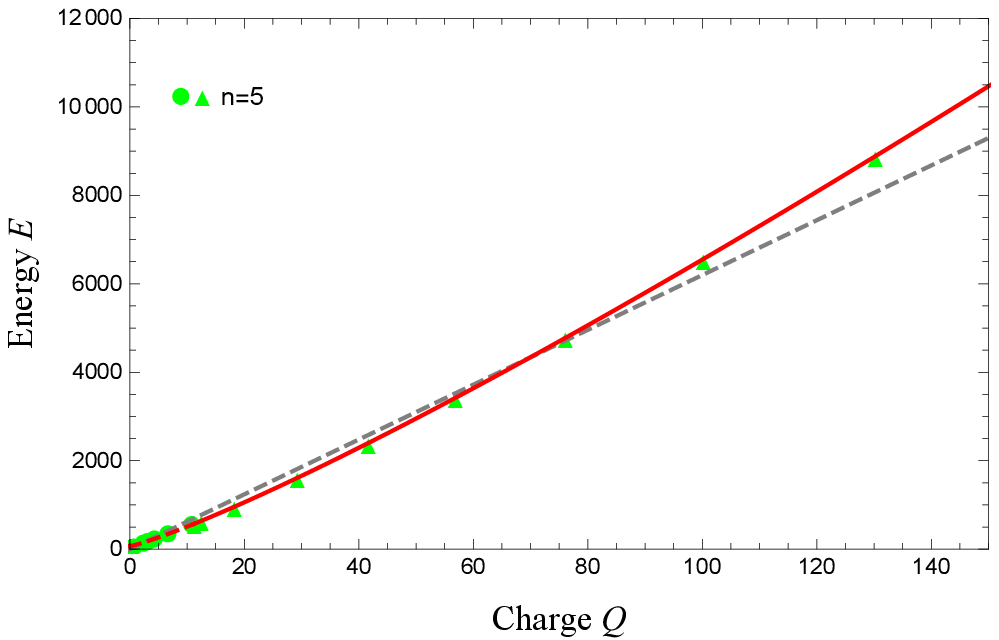}\\
    \vspace{5mm}

    \caption{\label{QEf} The relation between $E$ and $Q$. {\it Left}:~$\mathbb{C}P^{1}$. 
	{\it Right}:~$\mathbb{C}P^{11}$. The gray dashed lines indicate $E=\beta^{-5} Q$, where $\beta$ is the gradient of 
	$E^{-1/5}$-$Q^{-1/6}$ in Fig.\ref{Q6E5f}. 
	We numerically fit the data of the solutions in the second branch with $E =\eta Q^\alpha +\xi$, The resultant functions are 
	shown as the red line. For $\mathbb{C}P^1$, $\eta=1.12327 ,\xi=24.7417, \alpha=1.27389$. 
	For $\mathbb{C}P^{11}$ $\eta=30.0064, \xi=68.5474, \alpha=1.16711 \sim 7/6$.}
  \end{center}
	\end{figure*}
%%%%%%%%%%%%%%%%%%%%%%%%%%%%%%%%%%%%%%%%%%%%%%%%%%%%%%%%%%%%%%%%%%%%%%%%%%%%%%%%%%%%%%%%%%%%%%%%%%%%%%%%%%%%%%%%%

%%%%%%%%%%%%%%%%%%%%%%%%%%%%%%%%%%%%%%%%%%%%%%%%%%%%%%%%%%%%%%%%%%%%%%%%%%%%%%%%%%%%%%%%%%%%%%%%%%%%%%%%%%%%%%%%%
\begin{figure*}[t]
  \begin{center}
    \includegraphics[width=80mm]{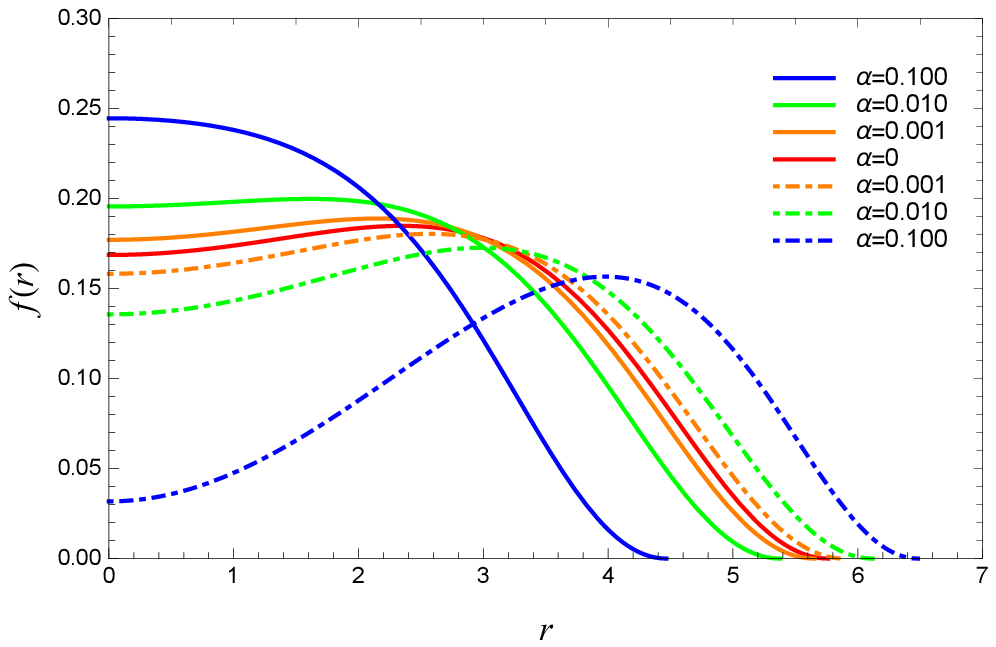}~~
    \includegraphics[width=80mm]{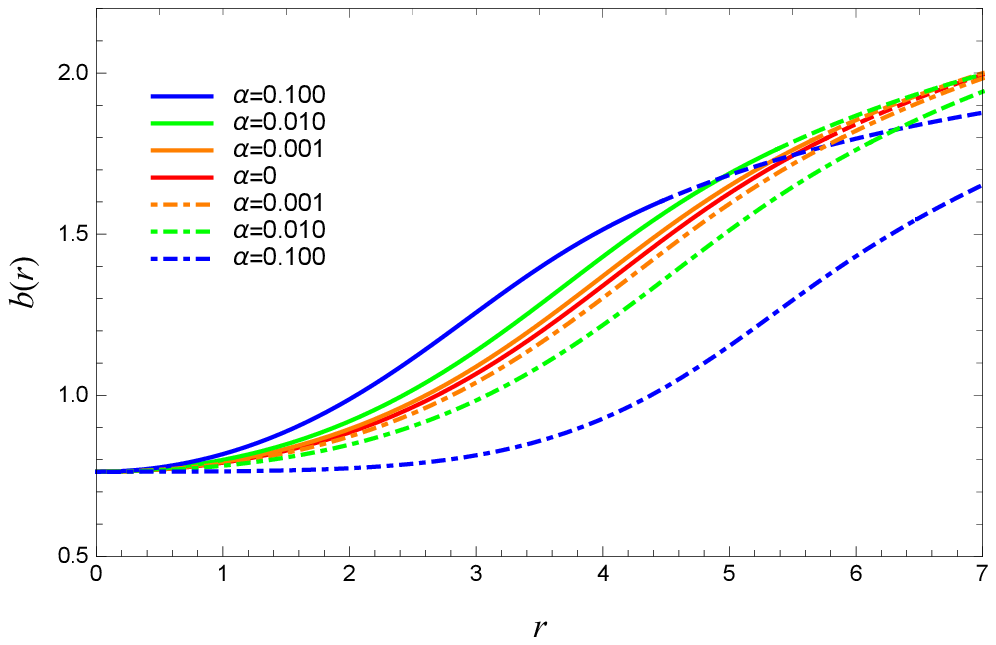}\\
    \vspace{5mm}

    \includegraphics[width=80mm]{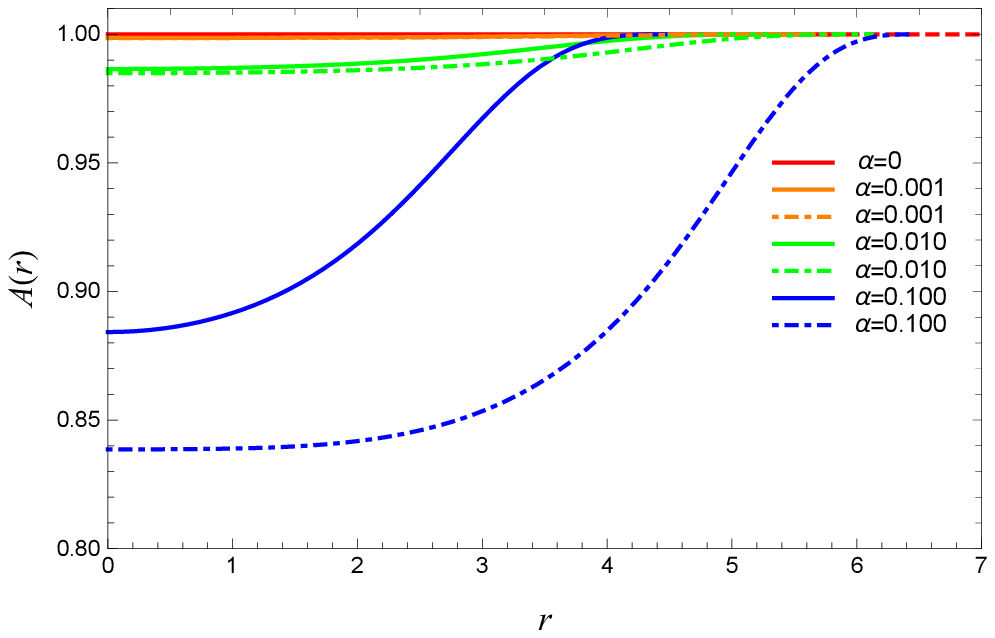}~~
    \includegraphics[width=80mm]{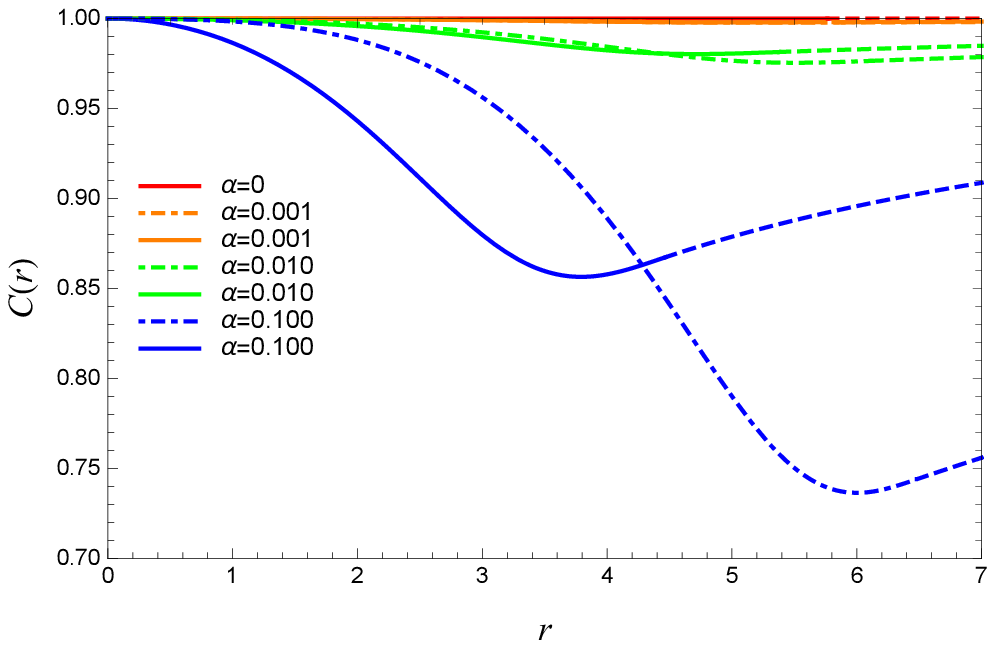}
    \caption{\label{CP1g} The gauged gravitating $Q$-ball solution for the $\mathbb{C}P^{1}$. 
The parameter $b_{0}$ is fixed as $b_{0}=0.7627$. 
{\it Top left}:~The matter profile function $f(r)$. 
{\it Top right}:~The gauge function $b(r)$. 
{\it Bottom left}:~The metric function $A(r)$.  
{\it Bottom Right}:~The metric function $C(r)$. 
Solutions of the first branch are plotted with bold line and the second branch are plotted with dot-dashed line. 
Solutions of the vacuum equations are depicted with the dashed line. 
Distinct curves correspond with different values of coupling constant $\alpha$.}
  \end{center}
	\end{figure*}
%%%%%%%%%%%%%%%%%%%%%%%%%%%%%%%%%%%%%%%%%%%%%%%%%%%%%%%%%%%%%%%%%%%%%%%%%%%%%%%%%%%%%%%%%%%%%%%%%%%%%%%%%%%%%%%%%

	\subsection{The Noether charge}
	The symmetry of the matter field is $SU(N) \otimes U(1)$. 
	Since it contains the $U(1)^N$ symmetry subgroup then following the reference \cite{Loginov:2016yeh}
	we consider a covariant derivative for the $\mathbb{C}P^N$ field 
	\begin{align}
	D_\mu u_i=\partial_\mu u_i-ieA_\mu {\mathcal Q}_{ij}u_j
	\end{align}
	where ${\mathcal Q}_{ij}$ is some real diagonal matrix: $Q_{ij}={\rm diag}(q_1,\cdots,q_N)$.
	The action (\ref{action}) with the covariant derivative is invariant under 
	following local $U(1)^N$ symmetry
	\begin{align}
	&A_\mu(x) \to A_\mu(x)+e^{-1}\partial_\mu\Lambda(x) \nonumber \\
	&u_i\to \exp[iq_i \Lambda(x)]u_i,~~~~i=1,\cdots,N\,.
	\label{gtransformation}
	\end{align}	
	The following Noether current is associated with the invariance of the action (\ref{action})  
	under transformations (\ref{gtransformation})
	\begin{align}
	J^{(i)}_\mu=-\frac{4M^2i}{\vartheta^4}\sum_{j=1}^N[u_i^* \Delta_{ij}^2 D_\mu u_j-D_\mu u_j^* \Delta_{ji}^2u_i]\,.
	\end{align}
	Using the ansatz (\ref{ansatzcpn}),(\ref{ansatzgauge}) we find the following form of the Noether currents
	\begin{align}
	J_t^{(m)}= \frac{(n-m)!}{(n+m)!}\frac{8bf^2}{(1+f^2)^2}\bigl(P^m_n(\cos\theta)\bigr)^2\,,
	\label{current0} \\
	J_\varphi^{(m)}=\frac{(n-m)!}{(n+m)!}\frac{8mf^2}{(1+f^2)^2}\bigl(P^m_n(\cos\theta)\bigr)^2
	\label{currentp}
	\end{align}
	and $J_r^{(m)}=J_\theta^{(m)}=0$ for $m=-n,-n+1,\cdots,n-1,n$.
	The conservation of currents is explicit after writing the continuity equation in the form
	\begin{align}
	\frac{1}{\sqrt{-g}}\partial_\mu (\sqrt{-g}g^{\mu\nu}J_\nu^{(m)})=\frac{1}{A^2C}\partial_t J_t^{(m)}
	+\frac{1}{r^2\sin^2\theta}\partial_\varphi J_\varphi^{(m)}=0
	\end{align}	
	Therefore, the corresponding Noether charge is
	\begin{align}
	Q^{(m)}&=\frac{1}{2}\int_{\mathbb{R}^3} d^3x \sqrt{-g} \frac{1}{A^2C}J^{(m)}_t(x) \nonumber \\
	&=\frac{16\pi}{2n+1}\int r^2dr\frac{bf^2}{AC(1+f^2)^2}\,.
	\end{align}
	Owing to our ansatz, the charge does not depend on index $m$, which means the symmetry of the solutions 
	is reduced to the $U(1)$. However, we shall keep the index for completeness. 
	
	The spatial components of the Noether currents 
	do not contribute to the charges; however, they can be used to introduce some auxiliary integrals~\cite{Klimas:2018ywv}
	\begin{align}
	q^{(m)}&=\frac{3}{2}\int d^3x\sqrt{-g}\frac{J_\varphi^{(m)}(x)}{r^2} \nonumber \\
	&=\frac{48\pi m}{2n+1}\int^\infty_0 dr \frac{Af^2}{1+f^2}\,.
	\end{align}
	
	As in the case of the gauged $Q$-balls~\cite{Lee:1988ag}, and also in the compactons~\cite{Arodz:2008nm}, 
	the total energy of our gauged model can be expressed using these Noether charges.  
	For the $Q$-shells, the function $f(r)$ vanishes for $r<R_{\rm in}$ and $r>R_{\rm out}$, so the Noether charge $Q^{(m)}$ is
	\begin{align}
	Q^{(m)}=\frac{16\pi}{2n+1}\int_{R_{\rm in}}^{R_{\rm out}} r^2dr\frac{bf^2}{AC(1+f^2)^2}\,.
	\end{align}
	The equation (\ref{eq:b}) is written in the compact form
	\begin{align}
	\kappa\biggl(r^2\frac{b'}{A}\biggr)'=\frac{8e^2r^2}{AC}\frac{bf^2}{(1+f^2)^2}\,.
	\label{eqcompact}
	\end{align}
	A single integration gives expression  	
	\begin{align}
	b'(r)=\frac{1}{r^2}\frac{8e^2A(r)}{\kappa}\int_0^r r'^2dr'\frac{b(r')f(r')^2}{A(r')C(r')(1+f(r')^2)^2}
	\end{align}
	which implies that the function $b(r)$ is a monotonically increasing function. 
	Therefore, for sufficiently large $r~(>R_{\rm out})$, $b'(r)= \bar{Q}/r^2~(\bar{Q}>0)$, where
	\begin{align}
	\bar{Q}\equiv \frac{8e^2A}{\kappa}\int_{R_{\rm in}}^{R_{\rm out}} r^2dr\frac{bf^2}{AC(1+f^2)^2}
	=\frac{e^2A(2n+1)}{2\pi\kappa}Q^{(m)}
	\label{Qbar}
	\end{align}
	where we have used the boundary condition: $A(\infty)\equiv 1$.
	Thus we obtain $b(r)$ for large $r$
	\begin{align}
	b(r)=\omega-\frac{\bar{Q}}{r}\,.
	\label{asymb}
	\end{align}
	$A_t$ behaves as $A_t=\bar{Q}/e r$ for large $r$ and therefore it can be interpreted as the 
	Coulomb potential of the spherically symmetric charge distribution in the compact region.
	Second term of the right-hand side of (\ref{energy}) can be evaluated by the partial integration 
	\begin{align}
	&\frac{1}{2}\int r^2dr\frac{\kappa b'^2}{Ae^2}=\frac{\kappa}{2}\biggl[r^2\frac{b'b}{Ae^2}\biggr]^\infty_0
	-\frac{\kappa}{2}\int dr b \biggl(r^2\frac{b'}{Ae^2}\biggr)' \nonumber \\
	&~~=\frac{\kappa}{2}\biggl(r^2\frac{b'b}{Ae^2}\biggr)\biggr|_{r\to\infty}
	-\frac{1}{2}\int dr\frac{8r^2}{AC}\frac{b^2f^2}{(1+f^2)^2}\,.
	\label{eq:sec}
	\end{align}
	where we have used (\ref{eqcompact}).
	The first term of the right hand side can be evaluated with (\ref{Qbar}) 
	and with the asymptotic behavior of $b$ (\ref{asymb})
	\begin{align}
	r^2\frac{b'b}{Ae^2}\biggr|_{r\to\infty}
	=r^2\frac{1}{A2e^2}\omega\biggl(\frac{\bar{Q}}{r^2}\biggr)
	=\omega\frac{(2n+1)}{2\pi\kappa}Q^{(m)}
	\end{align}
	As a result, we obtain the total energy for large $r$ of the form
	\begin{align}
	E=\sum_{m=-n}^n(\omega Q^{(m)}+mq^{(m)})
	+4\pi\int r^2drA\biggl(\frac{4Cf'^2}{(1+f^2)^2}+V\biggr)\,.
	\end{align}
	This form is similar of the non-gauged case~\cite{Klimas:2018ywv}.
	
	For the full understanding of the gauged $Q$-ball boson stars, we need to know $E$ as function of $Q$. 
	Only limited cases such like a thin-wall approximation of the model in a flat space-time might be possible, 
	but for the gravitating case, we have to rely on the numerical analysis.

%%%%%%%%%%%%%%%%%%%%%%%%%%%%%%%%%%%%%%%%%%%%%%%%%%%%%%%%%%%%%%%%%%%%%%%%%%%%%%%%%%%%%%%%%%%%%%%%%%%%%%%%%%%%%%%%%
\begin{figure*}[t]
  \begin{center}
    \includegraphics[width=80mm]{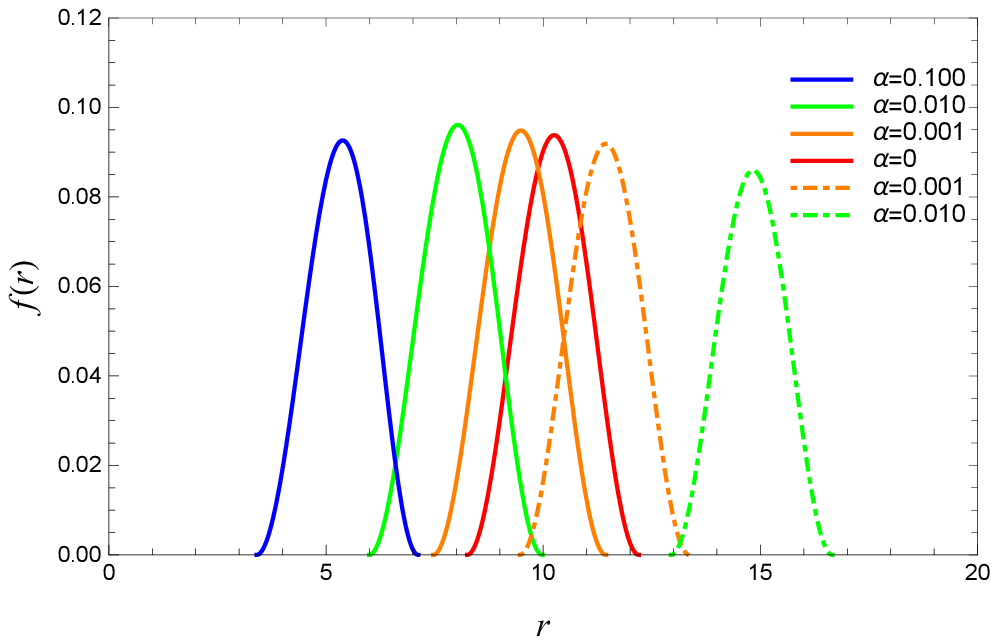}~~
    \includegraphics[width=80mm]{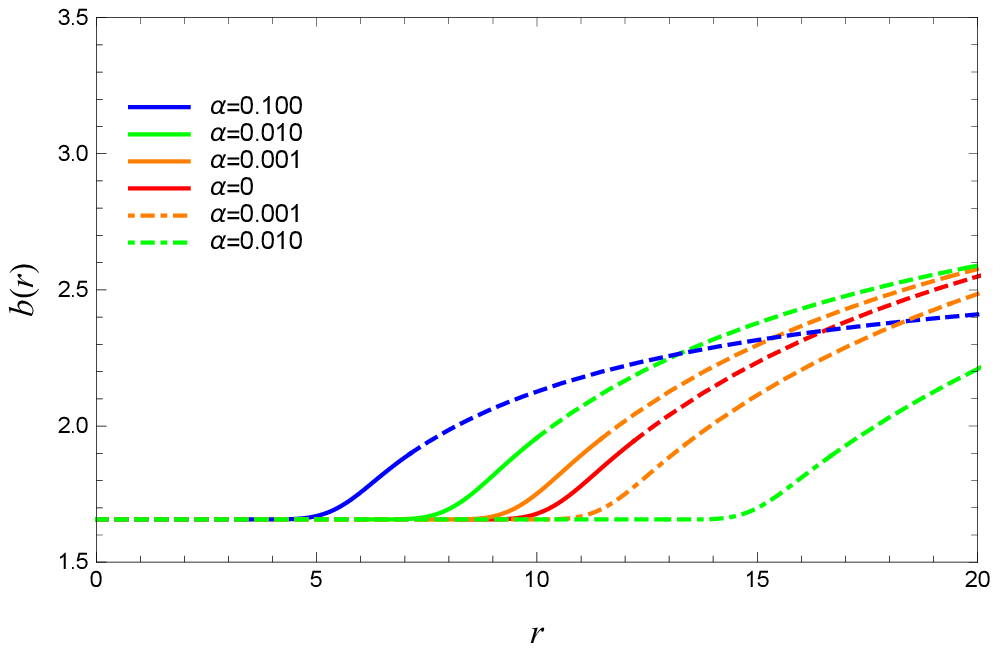}\\
    \vspace{5mm}

    \includegraphics[width=80mm]{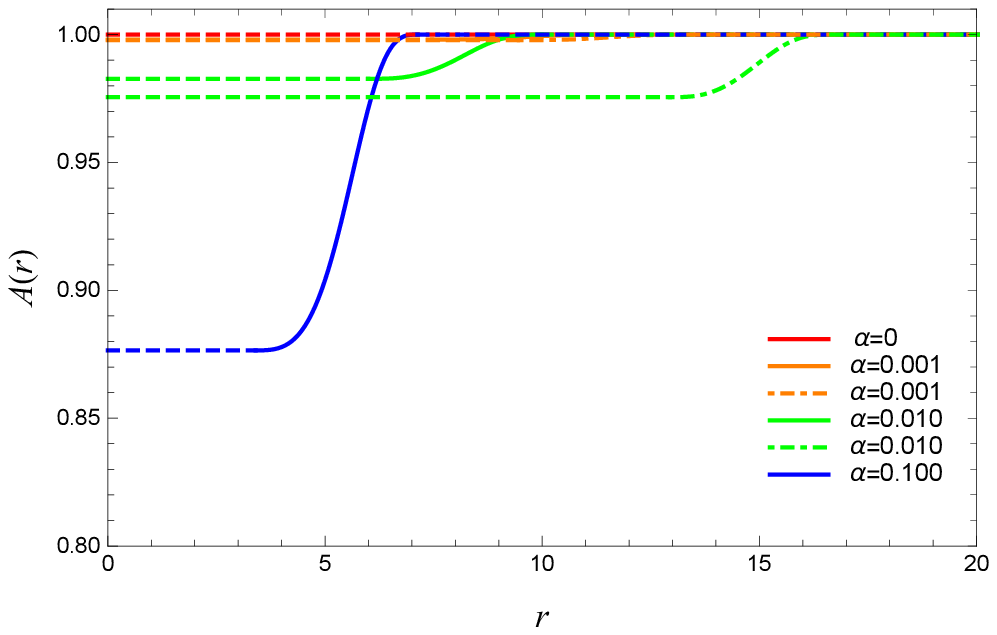}~~
    \includegraphics[width=80mm]{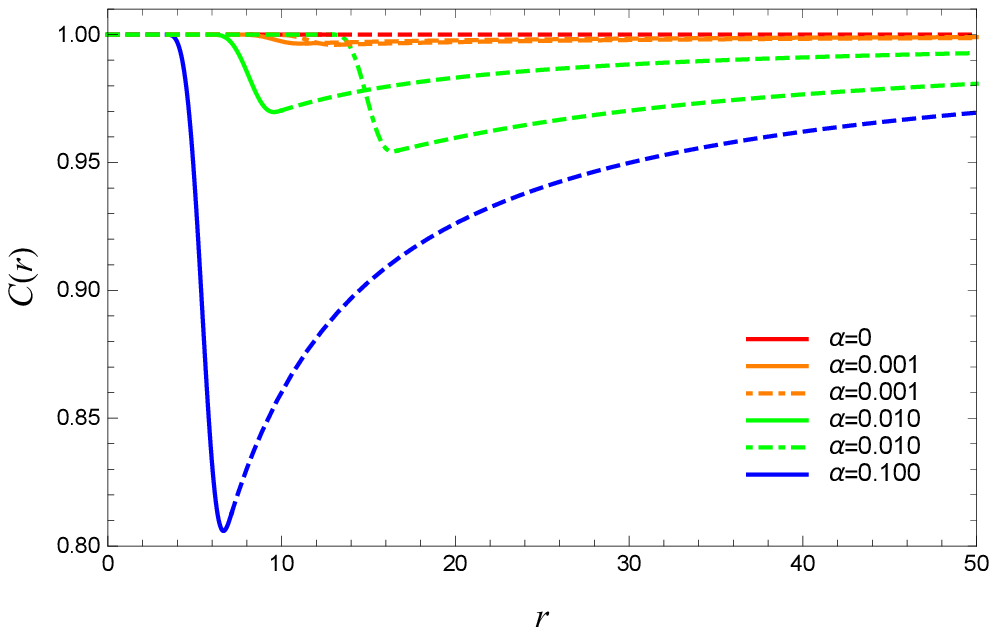}
    \caption{\label{CP11g} The gauged gravitating $Q$-shell solution for the $\mathbb{C}P^{11}$. 
The parameter $b_{0}$ is fixed as $b_{0}=1.657$. 
{\it Top left}:~The matter profile function $f(r)$. 
{\it Top right}:~The gauge function $b(r)$. 
{\it  Bottom left}:~The metric function $A(r)$. 
{\it  Bottom right}:~and the metric function $C(r)$. 
Solutions of the first branch are plotted with bold line and 
the second branch are plotted with dot-dashed line. 
Solutions of the vacuum equations are depicted with the dashed line. 
Distinct curves correspond with different values of coupling constant $\alpha$.}
  \end{center}
	\end{figure*}
%%%%%%%%%%%%%%%%%%%%%%%%%%%%%%%%%%%%%%%%%%%%%%%%%%%%%%%%%%%%%%%%%%%%%%%%%%%%%%%%%%%%%%%%%%%%%%%%%%%%%%%%%%%%%%%%%

\begin{figure*}[t]
  \begin{center}
    \includegraphics[width=80mm]{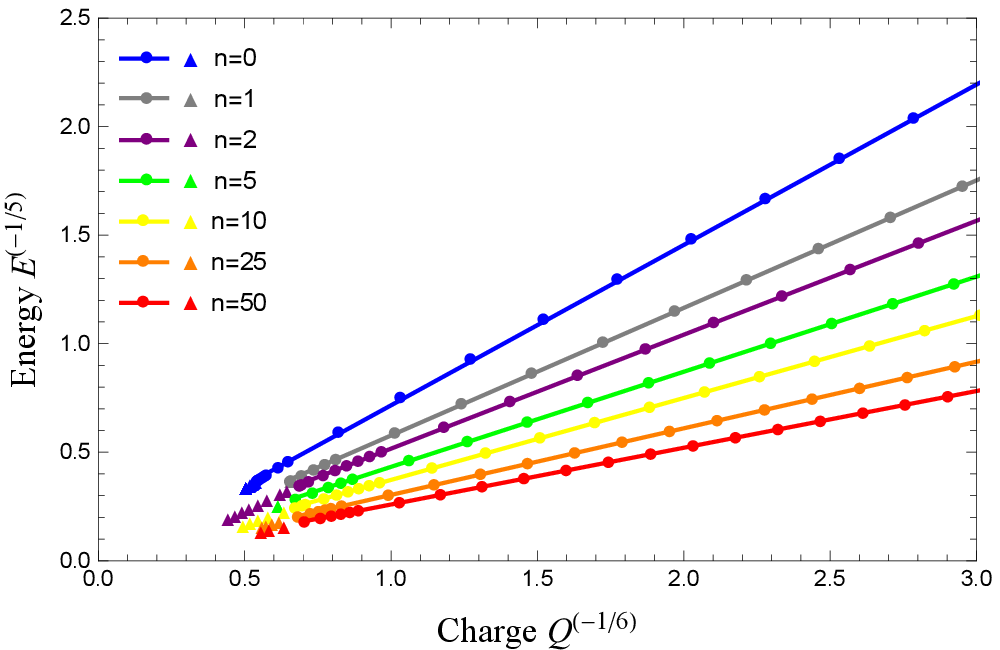}~~
    \includegraphics[width=80mm]{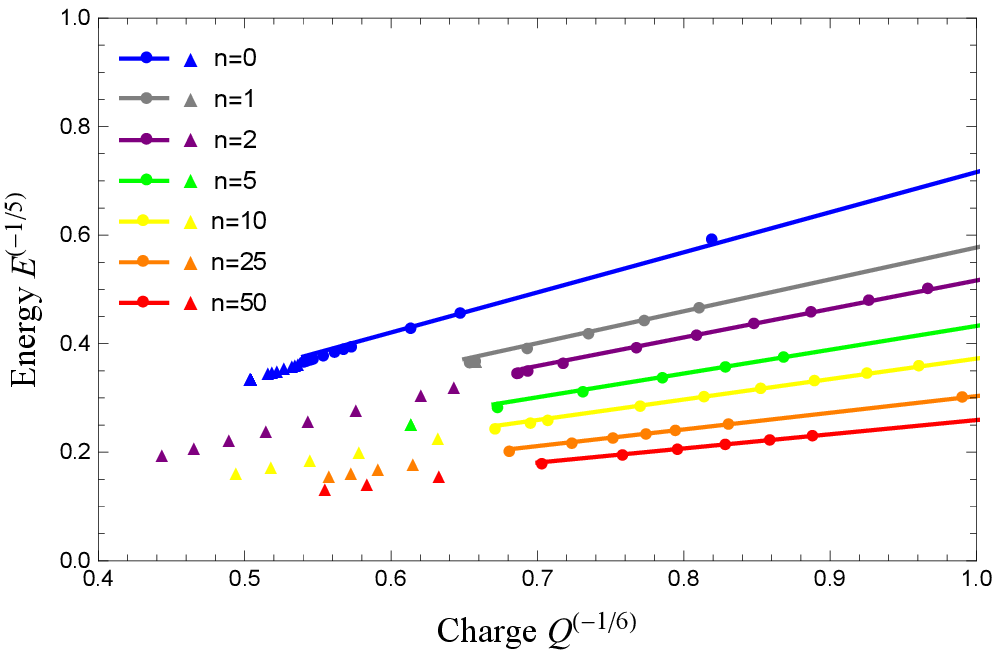}\\
    \vspace{5mm}
    \caption{\label{QEa2} {\it Left}:~The relation between $E^{-1/5}$ and $Q^{-1/6}$ for several gauged solutions. 
 {\it Right}: The same as the left one but the plot is enlarged in region of high $Q,E$.
The parameter $\alpha = 0.001$. 
The dots correspond to solutions with $\omega$ of the first branch. The triangles are of the second branch. }
  \end{center}
	\end{figure*}
%%%%%%%%%%%%%%%%%%%%%%%%%%%%%%%%%%%%%%%%%%%%%%%%%%%%%%%%%%%%%%%%%%%%%%%%%%%%%%%%%%%%%%%%%%%%%%%%%%%%%%%%%%%%%%%%%
%%%%%%%%%%%%%%%%%%%%%%%%%%%%%%%%%%%%%%%%%%%%%%%%%%%%%%%%%%%%%%%%%%%%%%%%%%%%%%%%%%%%%%%%%%%%%%%%%%%%%%%%%%%%%%%%%

\begin{figure*}[t]
  \begin{center}
    \includegraphics[width=80mm]{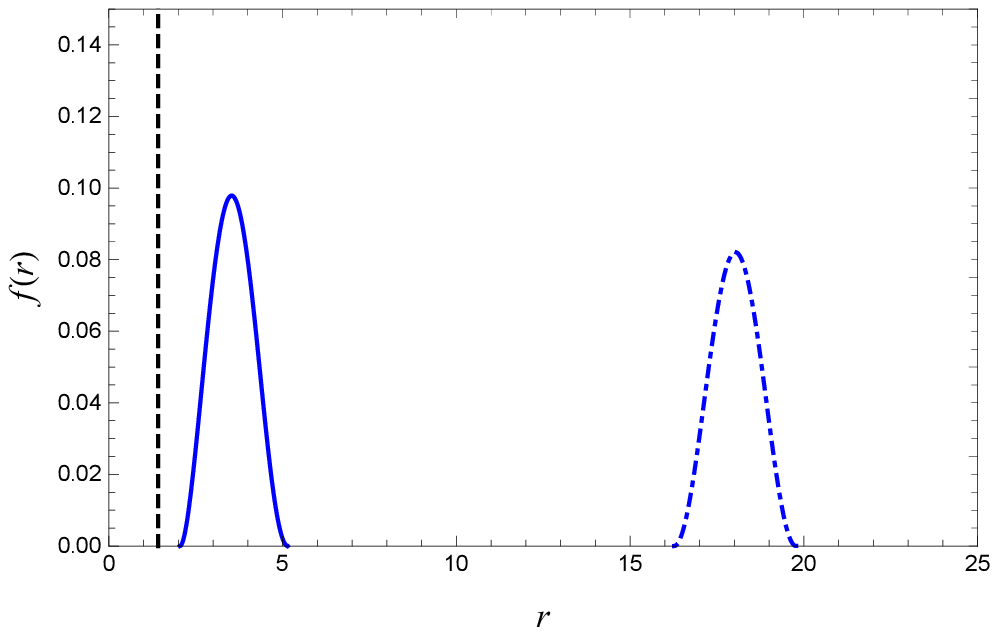}~~
    \includegraphics[width=80mm]{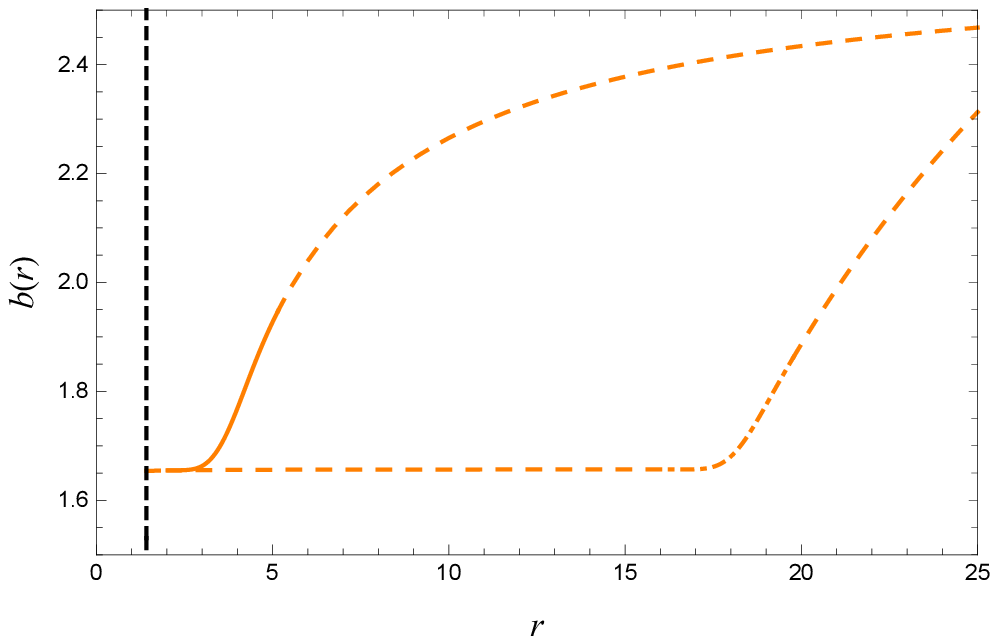}\\
    \vspace{5mm}
    \includegraphics[width=80mm]{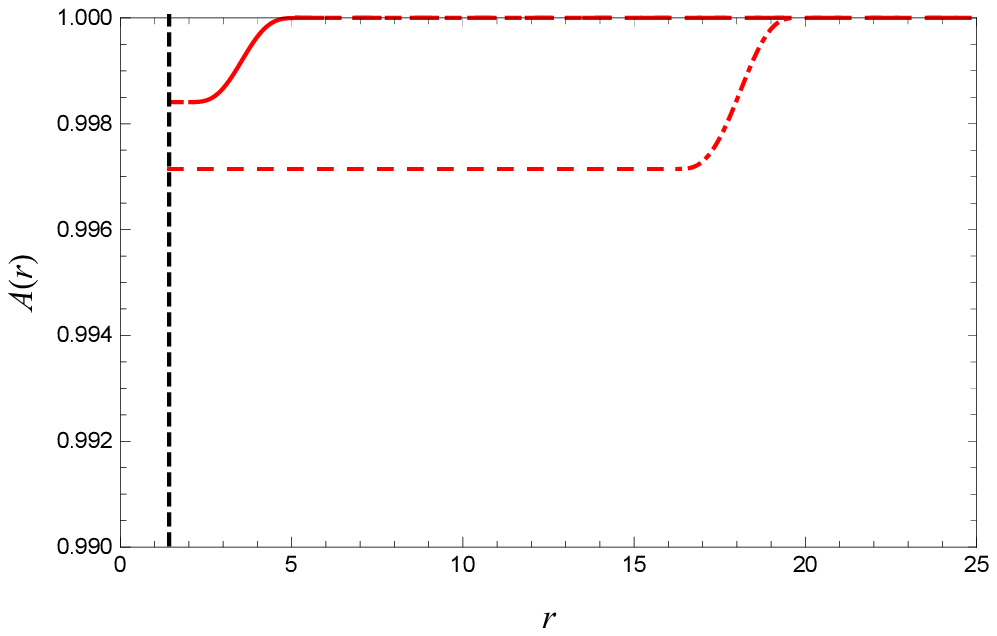}~~
    \includegraphics[width=80mm]{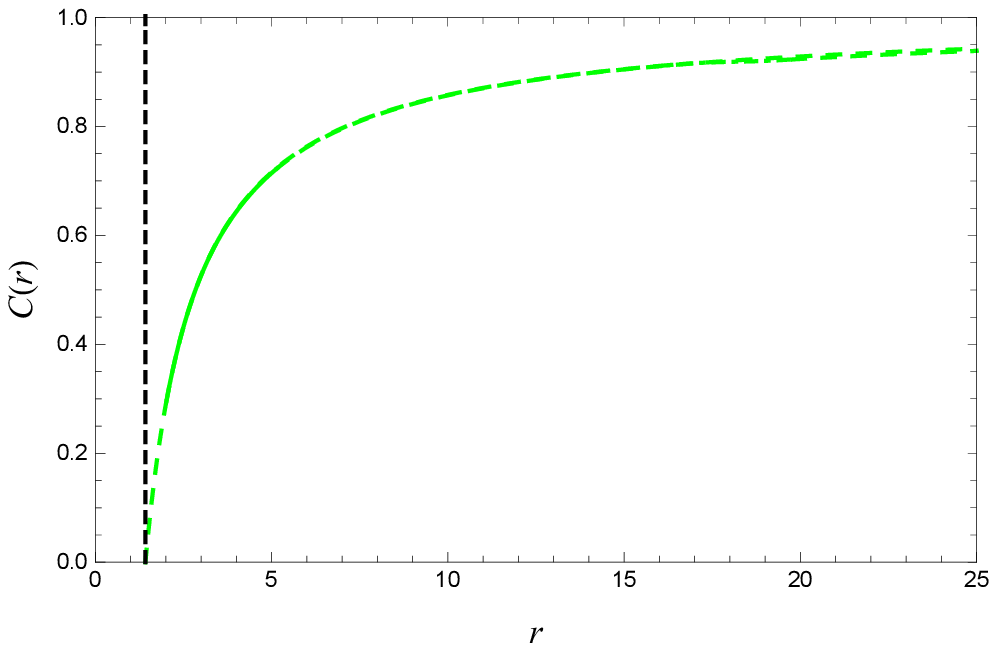}\\
    
    \caption{\label{CP11bh}  The harbor solution for $\mathbb{C}P^{11}$ for a charged black hole. 
The parameters $b_{0}=1.657$ and $\alpha=0.001$. 
{\it Top left}:~The matter profile function $f(r)$. 
{\it Top right}:~The gauge function $b(r)$. 
{\it Bottom left}:~The metric function $A(r)$.
{\it Bottom right}:~The metric function $C(r)$. 
Solutions of the first branch are plotted with bold line 
and the second branch are plotted with dot-dashed line. 
Solutions of the vacuum equations are depicted with the dotted line. 
The radius of the event horizon is chosen as $r_{\rm H}=1.420$.}
  \end{center}
	\end{figure*}
%%%%%%%%%%%%%%%%%%%%%%%%%%%%%%%%%%%%%%%%%%%%%%%%%%%%%%%%%%%%%%%%%%%%%%%%%%%%%%%%%%%%%%%%%%%%%%%%%%%%%%%%%%%%%%%%%

%%%%%%%%%%%%%%%%%%%%%%%%%%%%%%%%%%%%%%%%%%%%%%%%%%%%%%%%%%%%%%%%%%%%%%%%%%%%%%%%%%%%%%%%%%%%%%%%%%%%%%%%%%%%%%%%%
\begin{figure*}[t]
  \begin{center}
    \includegraphics[width=80mm]{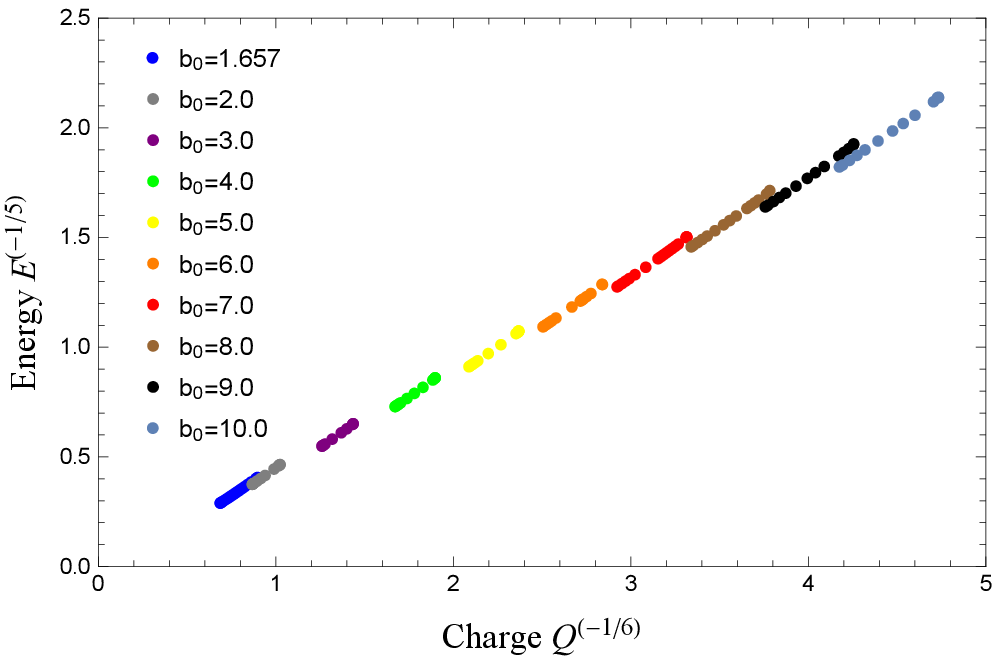}~~
    \includegraphics[width=80mm]{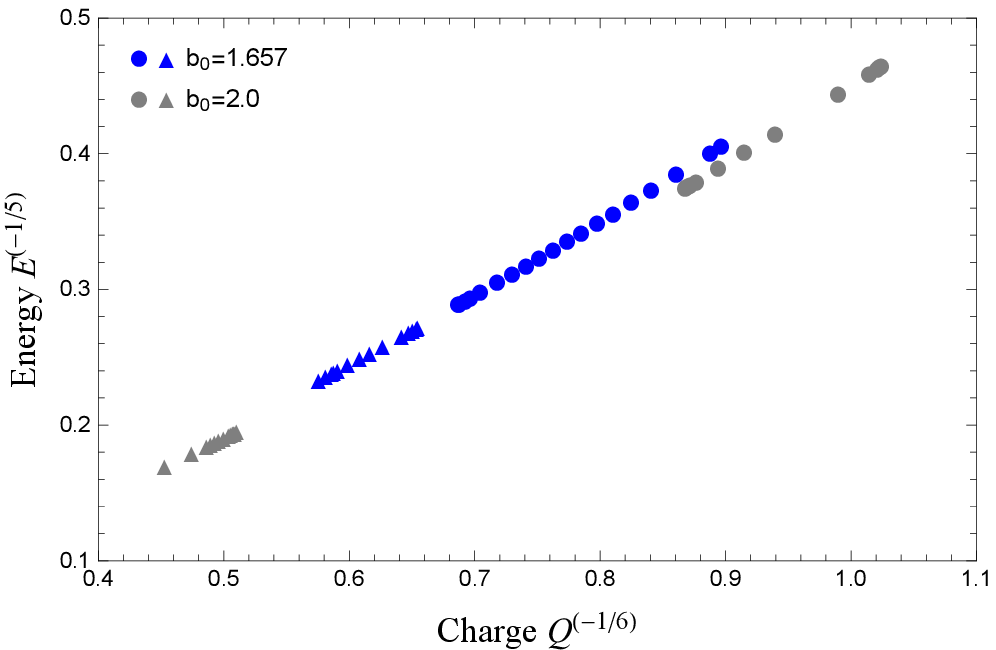}
    \vspace{5mm}
    
    \caption{\label{QEbh}  {\it Left}:~The relation between $E^{-1/5}$ and $Q^{-1/6}$ for harbor solutions of $\mathbb{C}P^{11}$.
The dots with the same color indicate the solutions which differ only by the value of a horizon radius $r_{\rm H}$. The dots 
corresponds the first branch.  
{\it Right}:~Same as the left but the plots for $b_{0} = 1.657, 2.0$ are enlarged. The dots correspond to the first branch 
and the triangles are the second branch. 
The parameters $\alpha=0.001$. }
  \end{center}
	\end{figure*}
%%%%%%%%%%%%%%%%%%%%%%%%%%%%%%%%%%%%%%%%%%%%%%%%%%%%%%%%%%%%%%%%%%%%%%%%%%%%%%%%%%%%%%%%%%%%%%%%%%%%%%%%%%%%%%%%%

%%%%%%%%%%%%%%%%%%%%%%%%%%%%%%%%%%%%%%%%%%%%%%%%%%%%%%%%%%%%%%%%%%%%%%%%%%%%%%%%%%%%%%%%%%%%%%%%%%%%%%%%%%%%%%%%%
\begin{figure*}[t]
  \begin{center}
    \includegraphics[width=80mm]{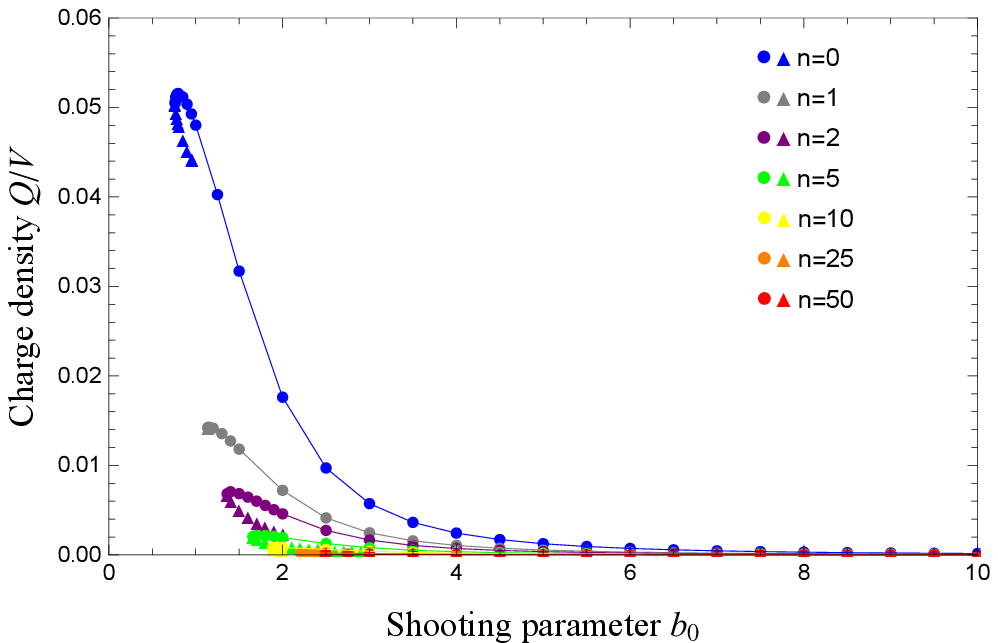}~~
    \includegraphics[width=80mm]{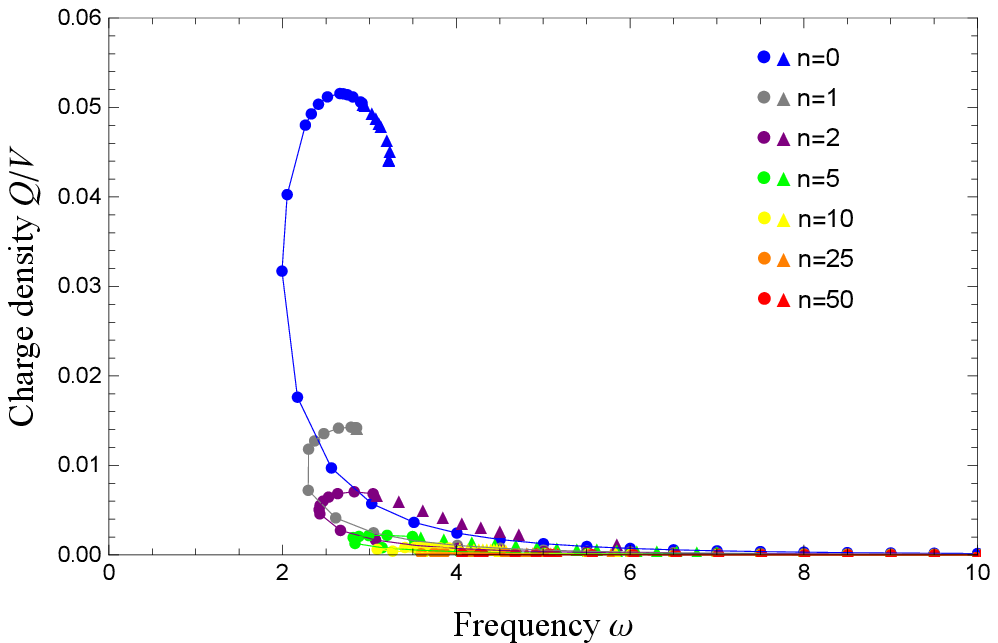}\\
	\vspace{1.0cm}
	
    \includegraphics[width=80mm]{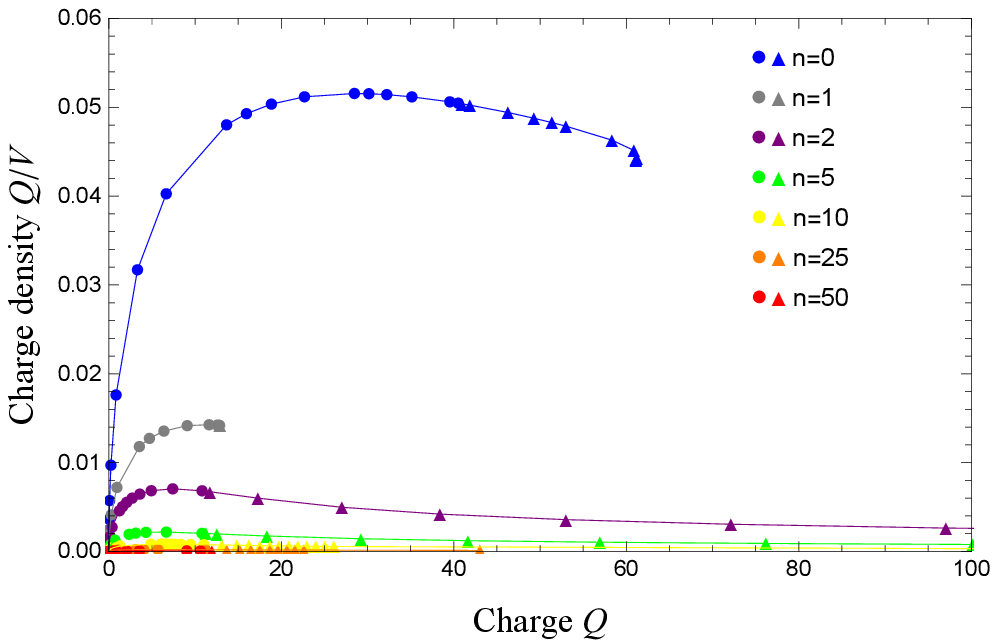}
    \vspace{5mm}
    
    \caption{\label{CDS}  The charge density: (The Noether charge $Q$)/(The volume of the $Q$-shell $V$) 
	for the several quantities of the model. 
	{\it Left top}:~For the shooting parameter $b_0$.
	{\it Right top}:~For the frequency $\omega$. {\it Bottom}:~the charge $Q$. 
	The dots correspond to the first branch and the triangles are the second branch.}
  \end{center}
	\end{figure*}
%%%%%%%%%%%%%%%%%%%%%%%%%%%%%%%%%%%%%%%%%%%%%%%%%%%%%%%%%%%%%%%%%%%%%%%%%%%%%%%%%%%%%%%%%%%%%%%%%%%%%%%%%%%%%%%%%

	\subsection{The boundary behavior of solutions}	

	We examine behavior of solutions at the boundary, 
	which means that we mainly look at the origin $r=0$ and the border(s) of the compacton. 
	First, we consider expansion at the origin and so the solution is represented by series
	\begin{align}
	&f(r)=\sum_{k=0}^\infty f_kr^k\,,~~~~
	b(r)=\sum_{k=0}^\infty b_kr^k\,, \nonumber \\
	&A(r)=\sum_{k=0}^\infty A_kr^k\,,~~~~
	C(r)=\sum_{k=-2}^\infty C_kr^k\,.
	\end{align}
	After substituting these expressions into equations (\ref{eq:N}), (\ref{eq:C}), (\ref{eq:f}),(\ref{eq:b}) 
	one requires vanishing of equations in all orders of expansion. It allows us to determinate the coefficients of expansion. 
	The form is given for each value of parameter $n$.
	For $n=0$ it reads
	\begin{align}
	&f(r)=f_0+\frac{1}{48}\biggl(\sqrt{1+f_0^2}-\frac{8f_0(1-f_0^2)b_0^2}{A_0^2(1+f_0^2)}\biggr)r^2+O(r^4)\,, \nonumber \\
	&b(r)=b_0 + \frac{4 e^2 b_0^2 f_0^2}{3(1+f_0^2)^2} r^2 + O(r^4)\,, \\
	&A(r)=A_0+\frac{2\alpha f_0^2b_0^2}{A_0(1+f_0^2)^2}r^2+O(r^4)\,,\label{expansion0} \\
	&C(r)=1-\frac{\alpha}{3}\biggl(\frac{f_0}{\sqrt{1+f_0}}+\frac{4f_0^2b_0^2}{A_0^2(1+f_0^2)^2}\biggr)r^2+O(r^4) \nonumber
	\end{align}
	where $f_0$, $ b_0$ and $A_0$ are free parameters.
	For $n=1$ we obtain
	\begin{align}
	&f(r)=f_1r+\frac{1}{32}r^2+\frac{1}{10}\biggl(2f_1^3(1+6\alpha)-\frac{f_1b_0^2}{A_0^2}\biggr)r^3+O(r^4)\,, \nonumber \\
	&b(r)=b_0 +\frac{2}{5}e^2 f_1^2 b_0 r^4 + O(r^5)\,, \\
	&A(r)=A_0+\alpha A_0f_1^2r^2+\frac{1}{6}\alpha A_0f_1r^3+O(r^4)\,,\label{expansion1} \\
	&C(r)=1-4\alpha f_1^2r^2 -\frac{\alpha f_1}{2}r^3+ O(r^4) \nonumber
	\end{align}
	with free parameters $f_1$, $b_0$ and $A_0$. 

	For $n\geqq 2$ we have no nontrivial solutions at the vicinity of the origin $r=0$, then
	the solution has to be identically zero. 
	In order to get nontrivial solution, we consider a possibility that the solution does not 
	vanish only inside the shell having radial support $r\in (R_{\rm in}, R_{\rm out})$. Solutions of this
	kind are called $Q$-shells. 
	We study expansion at the sphere with an inner or an outer radius. 
	Expansions at both borders of the compacton are very similar. 
	We impose the following boundary conditions at the compacton radius $r=R~(\equiv R_{\rm in},R_{\rm out})$
	\begin{align}
	f(R)=0,~~f'(R)=0,~~A(R)=1\,.
	\label{compacton}
	\end{align}
	The functions $f(r)$, $b(r)$, $A(r)$ and $C(r)$ are represented by series 
	\begin{align}
	&f(r)=\sum_{k=2}^\infty F_k(R-r)^k,~~b(r)=\sum_{k=0}^\infty B_k(R-r)^k,\nonumber \\
	&A(r)=\sum_{k=0}^\infty A_k(R-k)^k,~~ C(r)=\sum_{k=-2}^\infty C_k(R-r)^k.
	\end{align}
	First few terms have the form 
	\begin{align}
	&f(r)=\frac{R}{16C_0}(R-r)^2+\frac{R}{24C_0^2}(R-r)^3 +O((R-r)^4)\,, 
	\nonumber \\
	&b(r)=B_0 + B_1(R-r)-\frac{B_1}{R}(R-r)^2 + \frac{B_1}{3R^2}(R-r)^3 
	\nonumber \\
	&\hspace{5cm}+O((R-r)^4)\,, \nonumber \\
	&A(r)=A_0-\frac{\alpha R}{48C_0^2}(R-r)^3+O((R-r)^4)\,, \\
	&C(r)=C_0+\frac{1-C_0}{R}(R-r) 
	\nonumber \\
	&\hspace{1cm} +\biggl\{
	\biggl(C_0- 1 \biggr)\frac{1}{R_0^2}-\frac{5\alpha B_1^2}{4A_0^2e^2}
	\biggr\}(R-r)^2  \nonumber \\
	&\hspace{5cm} + O((R-r)^3) \,. \nonumber 
	\end{align}

	For electrically charged black hole solutions in the interior of the shell, we impose 
	the boundary conditions for the functions $b(r),C(r)$ at the inner radius $r=R_{\rm in}$
	\begin{align}
	&C(R_{\rm in})=1-\frac{2M_{\rm H}}{R_{\rm in}}+\frac{Q_{\rm H}^2}{R_{\rm in}^2}\,,
	\nonumber \\
	&b(R_{\rm in})=b_0-\frac{B_c}{R_{\rm in}},~~b'(R_{\rm in})=\frac{B_c}{R_{\rm in}^2}
	\end{align}
	where 
	\begin{align}
	M_{\rm H}=\frac{1}{2}\biggl(r_{\rm H}+\frac{Q_{\rm H}^2}{r_{\rm H}}\biggr),~~
	B_c=\frac{Q_{\rm H}A_0e\sqrt{2}}{\sqrt{\alpha}}
	\end{align}
	and $Q_{\rm H}$ is the horizon charge.

	\section{The solutions in the flat space-time}

We begin with the numerical analysis of $Q$-balls and $Q$-shells in flat space-time. 
We solve the coupled differential equations (\ref{eq:f}),(\ref{eq:b}) by a shooting method. 
According to (\ref{expansion0}) and (\ref{expansion1}), the solutions with $n=0,1$ are regular 
at the origin and then they are $Q$-ball type solutions. 
The solution of $n=0$ is plotted in Fig.\ref{CP1}.
The profile functions $f(r)$ are nonzero at the origin and monotonically approach zero 
of the compaction radius, which is similar behavior with the non-gauged solution discussed 
in \cite{Klimas:2018ywv}.
We have two independent shooting parameters $f_0, b_0$, and for the moment we fix $b_0$ 
and varies $f_0$ for finding solutions. 
The value of the $\omega$ is depicted via the asymptotic behavior of the numerical solution of $b(r)$, 
which obeys (\ref{asymb}). A notable feature of our solutions is that there is a branch, i.e., 
two independent solutions with equal values of shooting parameter, which exhibits a notable difference 
for varying $\omega$. 
In Fig.\ref{CP1} we present the typical behavior.
The solutions in the first branch (the bold lines) exhibit the lump like shape, i.e., the peaked at the origin. 
For smaller the shooting parameters $b_0$, 
the solutions $f(r)$ grow. On the other hand, the solutions in the second branch (the dot-dashed lines) 
have a dip, i.e., the peaks are located outside, not at the origin. They merge with $b_0=0.7627$. 
Interestingly, some of the solutions in the first branch ($b_0=0.7627$ and $0.80$) look interpolating between the branches. 
The solutions of the first branch and the second branch smoothly connect via these intermediate solutions.

As we posed in the previous section, the solutions are not regular at the origin for $n\geqq 2$, then 
leads to the $Q$-shells, i.e., the matter field is localized in the radial segment $r\in (R_{\rm in},R_{\rm out})$ 
and the gauge field $b(r)$ is of a constant at the interior region $r<R_{\rm in}$ 
and of the asymptotic solution (\ref{asymb}) at the exterior $r>R_{\rm out}$. 
Some typical solutions are shown in Fig.\ref{CP11}. 
Similar with the cases of the $Q$-balls, 
we obtain the solutions with branch which are merged at $b_0=1.657$.
The peak of the solutions moves outside as $\omega$ grows. Especially the second branch solutions rapidly expand 
as increasing $\omega$.  
In Fig.\ref{b0w}, we plot the relation between the shooting parameter $b_0$ and the corresponding frequency $\omega$
for $\mathbb{C}P^1$ and $\mathbb{C}P^{11}$.

For the stability of the $Q$-balls, we examine the energy-Noether charge scaling relation. 
In the studies for the global model, it was shown that the relation $E\sim Q^{5/6}$ are strictly holds for the 
solutions in flat space-time~\cite{Klimas:2017eft} and for several gravitating solutions~\cite{Klimas:2018ywv}. 
Thus, it is natural to investigate the relation between $E^{-1/5}$ and $Q^{-1/6}$ in the present model. 
In Fig.\ref{Q6E5f}, we plot the relation for $n=0,1,2,5,10,25,50$. 
The dots indicate the solutions with different frequency $\omega$. In each $n$, 
most of the points lie on the straight line with certain good accuracy. 
However, the data deviates from the linearity especially for larger $Q$ or $E$, and indicates for large $Q$
the energy scales different from $\sim Q^{5/6}$. Fig.\ref{QEf} are 
analysis of the relation between $Q$ and $E$ for $\mathbb{C}P^1$ and $\mathbb{C}P^{11}$, where the data
is depicted from result of Fig.\ref{Q6E5f}, which are compared to the dotted lines $E=\beta^{-1/5} Q$. 
The coefficient $\beta$ are extracted from the slope in Fig.\ref{Q6E5f},  i.e., $E^{-1/5}=\beta Q^{-1/6}$.
For the $Q$-ball, the energy is lower than the linear behavior.
The $Q$-shell solutions have the energy enhanced than the linearity. 
We study a power law of the scaling with numerically fitting the data for the second branch. 
For the $Q$-shell, it is $1.176711\sim 7/6$. 
This scaling is in good agreement with the result of $Q$-shell in the signum-Gordon model~\cite{Arodz:2008nm}.
The $Q$-ball solution has the power $\sim$ 1.2739, higher than $7/6$, which is a consequence of the boundary behavior 
at the origin. 
For growing the shooting parameter, the solutions of the second branch look closer to the shell shape, 
but not to become exact because we do not impose the compact support in the interior. 
Therefore, the compactness condition (\ref{compacton}) is essential for the scaling $Q^{7/6}$.

\section{The gravitating solutions}

The gravitating $Q$-balls and $Q$-shells are obtained by solving the differential equations (\ref{eq:N})-(\ref{eq:b})
with different $n$, We look at dependence of these solutions on the parameter $b_0, \alpha$. 
Here we present the numerical results for employing the value $b_0$ of the vicinity of the bifurcation point, 
i.e., the point where the branch merges.   
In Fig.\ref{CP1g}, we plot the solution for the $\mathbb{C}P^1$. 
In similarity to the flat case the profile function $f(r)$ has nonzero value
at the origin and reaches zero at the compacton radius. We also present the gauge function $b(r)$ and the 
metric functions $A(r),C(r)$. 
A notable feature of the solution is emergence of the second branch when varying the coupling constant $\alpha$. 
As $\alpha$ increases the solution of the first branch tends to shrink while the second one grows. 
Fig.\ref{CP11g} is the similar plots but for the $\mathbb{C}P^{11}$. The relation between $E^{-1/5}$ and $Q^{-1/6}$
of the gravitation solutions are presented in Fig.\ref{QEa2}.

We are able to consider the $Q$-shells with a massive body immersed in their center, which is referred as a harbor. 
In particular, there is a possibility of having this body as a Schwarzschild or a Reissner-Nordstr\"om type black-hole
~\cite{Kleihaus:2010ep,Kumar:2016sxx,Kumar:2014kna,Kumar:2015sia}. 
We set the event horizon in the interior part of the shell and then solve the equations from the event horizon 
to the outer region. 
In order to find the harbor solutions, we follow a few steps. 
We assume the solution of the vacuum Einstein equation in the region between the event horizon and 
the inner boundary of the shell $r\in [r_{\rm h},R_{\rm in}]$ because matter function vanishes 
in this region. 
A similar approach is applied outside the shell $r\in [R_{\rm out},\infty)$. 
Next, we solve the equations in the region $r\in (R_{\rm in}, R_{\rm out})$ and 
then smoothly connect the metric functions with such vacuum solutions at both boundaries.  
We present typical results in Fig.\ref{CP11bh} for $Q_{\rm H}=0.0001$. 

In Fig.\ref{QEbh}, we plot $E^{-1/5}$ - $Q^{-1/6}$ relation for the harbor solutions of $\mathbb{C}P^{11}$
with several values of $b_0$.

\section{Further discussion}

A salient feature of our $U(1)$ gauged model is a deviation from the energy-charge scaling $E\sim Q^{5/6}$ 
for large $Q$. 
Physically, we promptly guess that it is caused by the fact that the gauge field realizes the repulsive 
force between the constituents. The situation can be observed from  the behavior of the solutions in Fig.\ref{CP1}. 
At the low density, the solution behaves as a lump. It begins to deform at the intermediate region and  
the solutions of the second branch exhibit more delocalized, look like the shell structure, which apparently 
is effect of the repulsive electric force.  
In order to see the mechanism more qualitatively, we derive the charge density of our solutions. 
For simplicity, we employ the solutions of the flat space-time, but behavior of the gravitating solutions is 
quite similar. Thanks to the compactness of the solutions, we directly compute 
the volume of the $Q$-balls or $Q$-shells with the compacton radius $r=R_{\rm in},R_{\rm out}$. 
In Fig.\ref{CDS}, we present the charge density as a function of the shooting parameters $b_0$, the frequency $\omega$ 
and the charge $Q$. 
For the parameters $b_0$ or $\omega$, the density increases with the decrease the parameters from the large values.   
Approaching the maximum of the density, the solutions begin to deform and the density is relaxed. 
The solution moves on the second branch and the density turns to decreasing behavior. 
For the $Q$-shells, the energy scales as $Q^{7/6}$.  
For increasing the charge $Q$, the density also increases. 
The repulsive force by the electric interaction is effectively strong and 
then the solution deviates from $Q^{5/6}$ behavior.
After reaching the maximum, the solution relaxes for decreasing the density.

\section{Summary}

In this paper, we have considered $U(1)$ gauged $\mathbb{C}P^N$ nonlinear sigma model with a compact support
in flat space-time and also coupled with gravity. 
We have obtained the compact $Q$-ball and $Q$-shell solutions in the standard shooting method. 
The resulting self-gravitating regular solutions form boson-stars and boson-shells. 
For the compact $Q$-shell solutions we put the Schwarzschild like black holes in the interior 
and the exterior of the shell became the Reissner-Nordstr\"om space-time which may
be a contradiction of the no hair conjecture. 

In the $U(1)$ gauged model, the energy-charge scaling deviates from the corresponding global 
model, i.e., $E\sim Q^{5/6}$ for large $Q$. We discussed why the energy is enhanced for large $Q$ region 
in terms of the simple analysis of the charge density. For large $Q$, the charge density grows and then 
the repulsive force that originates in the electric interaction dominates and then the solution tends to be unstable. 

This paper is the first step for the construction of gravitating $Q$-balls (-shells) with non-Abelian 
symmetry $SU(2)\otimes U(1)$. $Q$-ball solutions with the symmetry $SU(2)\otimes U(1)$ will 
certainly be possible to exist. It is interesting because two different types of Noether 
charges corresponding to the symmetry have a crucial role in the stabilization of 
nontopological solitons. 
We shall report the results in our forthcoming paper. 
\vspace{0cm}

\vskip 0.5cm\noindent
\begin{center}
{\bf Acknowledgment}
\end{center}

	The authors would like to thank Pawe\l~Klimas for his careful reading of this manuscript 
	and also for many useful advices and comments. 
	We are also grateful for his kind hospitality of UFSC. 
	We appreciate Yuki Amari, Atsushi Nakamula, Kouichi Toda for valuable discussions. 
	S.Y. is grateful to Betti Hartmann of useful discussions and also the kind 
	hospitality of UPV/EHU where part of this work was done.
	S.Y. thanks the Yukawa Institute for Theoretical Physics at Kyoto University. 
	Discussions during the YITP workshop YITP-W-19-10 on "Strings and Fields 2019" 
	were useful to complete this work.
	N.S. is supported in part by JSPS KAKENHI Grant Number JP 16K01026 and B20K03278(1).

	\bibliography{cpncgdqshell}

\end{document}